\documentclass[twocolumn, untrackchanges]{aastex63}

\usepackage[T1]{fontenc}
\usepackage{graphicx}	
\usepackage{amsmath}	
\usepackage{amssymb}	
\usepackage{hyperref}
\usepackage{makecell}
\usepackage{color} 		

\shortauthors{Samuelsson et al.}
\shorttitle{Constraining llGRBs as UHECR sources}

\begin{document}
\label{firstpage}
\title{Constraining Low-luminosity Gamma-Ray Bursts as Ultra-high-energy Cosmic Ray Sources Using GRB 060218 as a Proxy}


\correspondingauthor{Filip Samuelsson}
\email{filipsam@kth.se}

\author[0000-0001-7414-5884]{Filip Samuelsson}
\affiliation{Department of Physics, KTH Royal Institute of Technology, \\
and The Oskar Klein Centre, SE-106 91 Stockholm, Sweden}

\author[0000-0003-4477-1846]{Damien B\'egu\'e}
\affiliation{Max-Planck-Institut f{\"u}r extraterrestrische Physik, Giessenbachstrasse, D-85748 Garching, Germany}

\author[0000-0002-9769-8016]{Felix Ryde}
\affiliation{Department of Physics, KTH Royal Institute of Technology, \\
and The Oskar Klein Centre, SE-106 91 Stockholm, Sweden}
 
\author[0000-0001-8667-0889]{Asaf Pe'er}
\affiliation{Department of Physics, Bar-Ilan University, Ramat-Gan 52900, Israel} 

\author[0000-0002-5358-5642]{Kohta Murase}
\affiliation{Department of Physics, The Pennsylvania State University, University Park, Pennsylvania 16802, USA}
\affiliation{Department of Astronomy \& Astrophysics, The Pennsylvania State University, University Park, Pennsylvania 16802, USA}
\affiliation{Center for Particle and Gravitational Astrophysics, The Pennsylvania State University, University Park, Pennsylvania 16802, USA}
\affiliation{Yukawa Institute for Theoretical Physics, Kyoto, Kyoto 606-8502 Japan}

\begin{abstract}
We study the connection between low-luminosity gamma-ray bursts (llGRBs) and ultra-high-energy cosmic rays (UHECRs) using the canonical low-luminosity GRB 060218 as a proxy. We focus on the consequential synchrotron emission from electrons that are coaccelerated in the UHECR acceleration region, comparing this emission to observations. Both the prompt and afterglow phases are considered. For the prompt phase, we assume the coaccelerated electrons are injected with a power law distribution instantaneously (without additional heating or reacceleration), which results in bright optical-UV emission in tension with observations. For the afterglow phase, we constrain the total kinetic energy of the blast wave by comparing electron thermal synchrotron radiation to available radio data at $\sim~3$ days. Considering mildly relativistic outflows with bulk Lorentz factor $\Gamma \gtrsim 2$ (slower trans-relativistic outflows are not treated), we find that the limited available energy does not allow for GRB 060218-like afterglows to be the main origin of UHECRs. This analysis independently constrains the prompt phase as a major UHECR source as well, given that the prompt energy budget is comparable to that of the afterglow kinetic energy. More generally, our study demonstrates that synchrotron emission from thermal electrons is a powerful diagnostic of the physics of mildly relativistic shocks.
\end{abstract}

\section{Introduction}
\label{Sec:Introduction}
Ultra-high-energy cosmic rays (UHECRs) are extraterrestrial particles with observed energies reaching from a few $10^{18}$ eV to above $10^{20}$ eV. Although they have been studied extensively, their origin still remains highly debated. Their energies are so high that they cannot be confined by the galactic magnetic field, indicating an extragalactic origin. It was early on suggested that gamma-ray bursts (GRBs) could be the accelerators of these particles \citep{Waxman1995, MilgromUsov1995, Vietri1995}. 
However, standard high-luminosity GRBs face several problems as UHECR accelerators. One of these is that the radiation field at the base of the jet is so intense that all heavy elements are disintegrated \citep[e.g.,][]{Wang2008, Murase2008, Horiuchi2012, Zhang2018}. This leads to a baryon outflow consisting mainly of protons, but both the Pierre Auger Telescope and the Telescope Array have seen indications that the UHECR composition consists of heavier nuclei toward the high-energy tail \citep{PierreAugerICRC2017, TelescopeArray2018}. 
A second problem is that UHECR acceleration requires the electrons in the GRB jet to be extremely fast-cooling in tension with observations, although this problem can be mitigated if the prompt emission and the UHECR acceleration occurs in separate regions of the ejecta \citep{Samuelsson2019}.
Furthermore, IceCube has put constraining upper limits on any coincident neutrinos from GRBs, which would be a clear sign of successful hadron acceleration \citep{IceCubeCollaboration2015GRBLimit}.

Low-luminosity GRBs (llGRBs) and trans-relativistic supernovae (TRSNe) have been studied by many authors as alternative candidates (\citealp[e.g.,][]{Murase2006, Wang2007, Murase2008, Chakraborti2011, Liu2011, Zhang2018, Boncioli2019, ZhangMurase2019}; \citealp[although see][]{Samuelsson2019,Anchordoqui2019arXiv}). These sources avoid the aforementioned disintegration problem present in the standard GRB-UHECR picture. The jet base radiation field is less intense due to their lower luminosity, so heavier particles can exist far out in the ejecta. 
The lower luminosity also makes them harder to detect, thus they could contribute significantly to the diffuse neutrino flux observed by IceCube without being electromagnetically visible. In the current work, we limit ourselves to studying llGRBs, by which we mean the relativistic or mildly relativistic part of the outflow that will start decelerating within the first few days. The possibility of UHECR acceleration at TRSNe associated with llGRBs will be discussed in Section \ref{Sec:DiscussionAfterglow}.

In this paper, we use GRB 060218 as a proxy to constrain llGRBs as UHECR sources. GRB 060218 is often considered the canonical llGRB. The existing data, ranging from radio to gamma-rays, is exceptionally good and it was monitored up to several months after the explosion. Due to the few identifications to date, it is difficult to say whether GRB 060218 is representative of the llGRB sample or not \citep{Sun2015}. This will become clear given enough future detections. If it is representative, then the results presented in this paper can be applied to the whole llGRB population (we argue in Section \ref{Sec:Comparison} that an extrapolation of the result to the larger population based on a single event might not be unfounded in this case). Else, the analysis can be remade with more typical parameters to see if any of the conclusions change. If they do, the results of this paper should be viewed as valid for GRB 060218 alone. 
We use GRB 060218 throughout the text to refer to the fastest part of the ejecta, responsible for the prompt X-ray emission and the radio emission in the first few days.

The methodology in this paper will follow closely that of \citet{Samuelsson2019}; we assume that electrons are also accelerated in the UHECR acceleration region. We then calculate the synchrotron emission from these electrons and compare it with observations. This is an approach that can efficiently constrain the acceleration site. 
Furthermore, the electron synchrotron spectrum carries additional information, in complement to the neutrino and hadronic gamma-ray signals created in the interactions between the cosmic rays and the ambient photon field. It should therefore always be considered in multi-messenger modeling of UHECR sources, which is not the case in most of the recent studies on the topic.
The difference of this work as compared to \citet{Samuelsson2019} is that in the current paper, we do not only consider possible acceleration of UHECRs in the prompt phase but also in the afterglow phase. Additionally, as we apply the method to a specific burst, we do not have to rely on generic flux limits for the observations but can resort directly to measurements. This makes the results presented here much more constraining, as the flux limits used in \citet{Samuelsson2019} were conservative.

The paper is organized as follows. We start by presenting GRB 060218 in Section \ref{Sec:GRB060218}. The prompt phase is considered in Section \ref{Sec:Prompt}, starting with the methodology in Subsection \ref{Sec:Methodology} followed by the results in Subsection \ref{Sec:Results} for fiducial parameters (\ref{Sec:Fiducial}) and for optimistic parameters (\ref{Sec:Optimistic}). We discuss the results for the prompt phase in Subsection \ref{Sec:DiscussionPrompt}. We go on to evaluate the afterglow phase in Section \ref{Sec:Afterglow}, presenting the methodology in Subsection \ref{Sec:MethodologyAG} and the results and discussion in Subsections \ref{Sec:ResultsAG} and \ref{Sec:DiscussionAfterglow}, respectively. Finally, we conclude in Section \ref{Sec:Conclusion}.

\section{GRB 060218}
\label{Sec:GRB060218}
\subsection{Observations and parameters}
GRB 060218 was first detected with the BAT instrument on board the Swift satellite. The closeness of the event and its association with the supernova SN 2006aj meant that it was extensively observed and studied \citep{Campana2006, Mazzali2006, Sollerman2006, Ferrero2006, Mirabal2006, Soderberg2006Nature, Pian2006, Fan2006, Ghisellini2007, Toma2007, Waxman2007, IrwinChevalier2016, Emery2019}. 

The prompt emission of this burst shows a single-peaked, very smooth light curve with an exceptionally long duration, $T_{90} = 2100 \pm 100$ s.
It was monitored with both the Swift XRT and UVOT telescopes during the prompt phase. Spectroscopic observation of the optical afterglow placed it at a redshift $z = 0.033$ \citep{GCN-4792, Pian2006}. It has a $\nu F_\nu$-peak in X-rays at around $E_\textrm{peak}\sim 5$ keV and an isotropically equivalent gamma-ray energy of $E_\gamma = (6.2 \pm 0.3) \times10^{49}$ erg, extrapolated to between $1-10^{4}$ keV \citep{Campana2006}. 
For a $T_{90}$ of 2100 s, this corresponds to an average radiation luminosity of $L_\gamma \sim 3\times10^{46}$ erg s$^{-1}$. 
\citet{Murase2006} were early in examining its possible connection to neutrino production and UHECR acceleration, arguing that the neutrino background from llGRBs could be comparable to that of high-luminosity GRBs. 

There are many theories as to what caused the prompt X-rays in GRB 060218. 
For our current analysis, the radiative origin of the prompt emission is of no importance. By not assuming that the observed prompt X-rays come from the UHECR acceleration site, we allow for a two-zone model, in which the prompt emission and the UHECR acceleration can occur in different parts of the ejecta. Indeed, we study the possibility of UHECR acceleration in the afterglow phase in Section \ref{Sec:Afterglow}. The only thing we are interested in is the unavoidable synchrotron emission from the coaccelerated electrons at the UHECR acceleration site. This emission should be consistent with (or lower than) the X-ray and optical observations for the prompt phase, and the radio observations for the afterglow phase.

When comparing the radiation from the coaccelerated electrons to observations, we have to account for extinction along the line of sight. Monitoring of the optical afterglow in GRB 060218 determined the Galactic reddening to $E(B - V) \sim 0.13$ mag and the host galaxy reddening to $E(B - V) \sim 0.04$ mag. For a ratio of total-to-selective extinction $R_V = 3.1$, this translates into into an extinction of $A_V \sim 0.39$ mag and $A_V \sim 0.13$ mag for our Galaxy and the host galaxy respectively \citep{GCN-4863, Ferrero2006, Pian2006}. These values of the reddening are the ones used for the spectrum in Figure 3 in \citet{Ghisellini2007}, which gives $F_{\nu_\textrm{opt}}^\textrm{obs} = 5.5\times 10^{-27}$ erg cm$^{-2}$ s$^{-1}$ Hz$^{-1}$ (0.55 mJy) and $F_{\nu_\textrm{X}}^\textrm{obs} = 10^{-27}$ erg cm$^{-2}$ s$^{-1}$ Hz$^{-1}$ (0.1 mJy) for the prompt, deabsorbed optical-UV and X-ray fluxes respectively. 
The subscripts indicate the frequencies at which we evaluate the flux. We use $h\nu_\textrm{opt} = 3~\textrm{eV}$ ($410~$nm) and $h\nu_\textrm{X} = 5~\textrm{keV}$ (observed frequencies), where $h$ is Planck's constant.
From afterglow observations, one can get an estimate of the initial bulk Lorentz factor of the outflow $\Gamma$. The radial distance from the progenitor of the radio emission five days after the trigger is estimated to be $r \sim 3 \times 10^{16}$ cm \citep{Soderberg2006Nature}. Assuming that the prompt emission radius is much smaller than the radius of the first afterglow light emission $r_\textrm{ag}$, one gets from the equation of motion $t_\textrm{ag} \lesssim r_\textrm{ag}/v - r_\textrm{ag}/c$, where $t_\textrm{ag}$ is the time from trigger to the onset of the afterglow, $v$ is the bulk velocity of the outflow, and $c$ is the speed of light in vacuum. From this, one obtains the upper limit
\begin{equation}\label{eq:Gamma_UL}
    \Gamma \lesssim \left[1 - \left(\frac{1}{t_\textrm{ag} c/r_\textrm{ag} + 1}\right)^2\right]^{-1/2}.
\end{equation}
With $t_\textrm{ag} > 2000$ s and $r_\textrm{ag} < 3\times 10^{16}$ cm, we obtain $\Gamma < 16$. While this value of $\Gamma$ is much lower than those for canonical GRBs, it is consistent with estimates from radio afterglow monitoring of llGRBs that indicate only mildly relativistic outflows with $\Gamma \gtrsim 2$ after a few days \citep{Soderberg2006Nature}. 

The upper limit on $\Gamma$ given in Equation \eqref{eq:Gamma_UL} is not appropriate if the engine duration is longer than the deceleration time. For instance, the prompt emission and early afterglow could have been caused by different parts of the ejecta, or the onset of the afterglow could have occurred earlier if it was initially outshone by the prompt emission. A mildly relativistic ejecta with $\Gamma \lesssim 10$ is in agreement with independent estimates from other authors, such as \citet{Campana2006, Soderberg2006Nature, Fan2006, Ghisellini2007, Toma2007, Waxman2007}. For completeness, we show results for $\Gamma = 3$, 10, and 30 in this work. 

Several authors have modeled the late-time emission of GRB 060218 as radiation from the forward shock of a blast wave propagating into the circumburst medium. They commonly find that a constant circumburst medium number density of $n_\textrm{cbm} = 100$ cm$^{-3}$ matches the data well \citep{Soderberg2006Nature,Fan2006,Toma2007}. A wind-like circumburst medium is disfavored by the temporal decay of the late-time 8.46 GHz light curve \citep{Fan2006,Toma2007}. 
They also find a rather low isotropically equivalent blast wave energy $E_\textrm{k}$ of $\sim 10^{48}\textrm{--}10^{50}$ erg. The radio data suggest a break in the spectrum at $\sim 4$ GHz at 5 days \citep{Soderberg2006Nature}. Due to the steep slope below the break, this is interpreted as the synchrotron self-absorption frequency \citep{Soderberg2006Nature,Fan2006,Toma2007}. 

\subsection{Requirement on the burst energy}\label{Sec:Energetics}
In this paper, we will use a total isotropically equivalent energy of $E_\textrm{tot} = 10^{51}$ erg for GRB 060218, higher than suggested by \citet{Soderberg2006Nature,Fan2006,Toma2007}. This value of the total energy translates to an average isotropically equivalent total luminosity of $L_\textrm{tot} = 4.8\times10^{47}$ erg s$^{-1}$. The reason for the larger value is that for llGRBs to be the main sources of UHECRs, the energy of the burst has to be sufficiently large to supply the observed UHECR flux at Earth. 

The UHECR energy injection rate is $E (dQ_\textrm{UHECR}/dE) \sim 10^{44}~$erg$~$Mpc$^{-3}~$yr$^{-1}$ with some uncertainty \citep{Waxman1995ApJ, Katz2009, MuraseTakami2009, Zhang2018}. \citet{AugerCollaboration2020} recently found the energy injection rate to be $\approx 6 \times 10^{44}~$erg$~$Mpc$^{-3}~$yr$^{-1}$ above $5\times10^{18}~$eV, although this result assumes no source density evolution with redshift, which in the case of GRBs corresponds to a factor of 3--5 difference. The apparent local event rate of llGRBs $R_{\rm LL,app}$ is estimated to be $10^2\textrm{--}10^3$ Gpc$^{-3}$ yr$^{-1}$ \citep{Soderberg2006Nature, Pian2006, Toma2007, Liang2007, MuraseTakami2009, Virgili2009, Sun2015}. Thus, on average, every event needs to release at least an isotropic equivalent energy of $Q_\textrm{UHECR}/R_{\rm LL,app} = 10^{50}\, Q_{\textrm{UHECR},44}R_{{\rm LL,app}, 3}^{-1}$ erg in UHECRs. Assuming a fraction $\xi_{\rm UHECR}$ of the total energy escapes as UHECRs, the necessary total isotropically equivalent energy is 
\begin{equation}\label{eq:E_tot}
\begin{split}
	E_\textrm{tot} &= \frac{Q_\textrm{UHECR}}{\xi_\textrm{UHECR} R_\textrm{LL,app}}\\[2.5mm]
	&= 10^{51} ~ Q_{\textrm{UHECR},44} ~ \xi_{\textrm{UHECR},-1}^{-1} ~ R_{\textrm{LL,app}, 3}^{-1} \text{ erg}.
\end{split}
\end{equation}
The choice $\xi_{\rm UHECR} = 0.1$ is conservative, as it includes the fraction of energy given to UHECRs (i.e., only protons and nuclei with $E > 3\times10^{18}~$eV), as well as the fraction of UHECRs that successfully escapes the system \citep{Wang2007, Murase2008, Chakraborti2011, ZhangMurase2019}. Furthermore, $R_\textrm{LL,app}$ is chosen at the optimistic end of its uncertainty range. Therefore, $10^{51}~$erg is really the minimum energy required. If $E_\textrm{tot}$ is increased, then all constraints presented here would become stronger. Conversely, if $E_\textrm{tot}$ is allowed to decrease due to an updated estimate putting the apparent rate $R_{\rm LL,app} > 10^3$ Gpc$^{-3}$ yr$^{-1}$, then our constraints would become less severe. Current studies suggests that $R_\textrm{LL,app} \sim ~\textrm{few}~100$ at most \citep{Sun2015}, but the error bars are large and future detections are needed to get a more precise estimate.




\subsection{GRB 060218 compared to other llGRBs}\label{Sec:Comparison}
Although GRB 060218 is extraordinary when looking at the sample of long GRBs, it is not as peculiar when compared to the handful of other detected llGRBs. 
A single-peaked, smooth light curve seems common for llGRBs \citep{Kaneko2007,NakarSari2012}, as well as a long duration and a soft peak \citep{Sun2015}. For instance, the low-luminosity GRB 100316D had an exceptionally long duration of at least 1300 s and a peak energy of $\sim 30$ keV \citep{Starling2011}. A low peak energy of $< 5$ keV was also found in the low-luminosity X-ray flash 020903 \citep{Sakamoto2004}. However, the llGRBs 980425, 031203, and 171205A were harder and shorter with $E_\textrm{peak}\gtrsim 125$ keV and a $T_{90}$ between 23 and 195 s \citep{Galama1998, Sazonov2004, DElia2018}, which shows that the sample is not uniform.  These differences could be diverse expressions of the same phenomena \citep[see e.g.,][for a common explanation of all llGRBs as shock break-outs in or out of thermal equilibrium]{NakarSari2012} or they could hint at bimodality within the llGRB sample. 

In this study, what is most important are the prompt optical fluxes and afterglow radio fluxes. Therefore, it is of specific interest to see whether other llGRBs are similar to GRB 060218 in these regards. 
Prompt optical fluxes are unfortunately rare. Apart from GRB 060218, only GRB 100316D and GRB 171205A have reports on the optical flux during the prompt emission phase. For GRB 100316D, UVOT 3$\sigma$ upper limits puts the $u$-filter magnitude at $>19.3$ \citep{Starling2011}. This corresponds to a deabsorbed flux of $< 8\times 10^{-28}$ erg cm$^{-2}$ s$^{-1}$ Hz$^{-1}$ (< 0.08 mJy) \citep{Fan2011}. With a redshift of $z = 0.0591$ \citep{Starling2011}, GRB 100316D was at least a factor of $\sim 2$ dimmer in prompt optical emission compared to GRB 060218. UVOT observation in the $u$-filter of GRB 171205A similarly puts the deabsorbed prompt optical flux at $8\times 10^{-28}$ erg cm$^{-2}$ s$^{-1}$ Hz$^{-1}$ (0.08 mJy) \citep{DElia2018}. With a redshift of $z=0.037$ \citep{DElia2018}, the prompt optical luminosity of GRB 171205A was $\sim 6$ times lower than that of GRB 060218. 

The $8.5~$GHz radio afterglows of llGRBs are universally less luminous than those of high-luminosity GRBs \citep{Margutti2013}. The luminosity variation at $8.5~$GHz for the llGRB sample in \citet{Margutti2013}, is roughly an order of magnitude at early times. A 3$\sigma$ upper limit at 1.8 days shows that the $8.5~$GHz radio luminosity of GRB 100316D was lower than that of GRB 060218 at the corresponding time. GRB 980425 and GRB 031203 were $\sim 2$ and $\sim 7$ times more luminous at $8.5~$GHz respectively, compared to GRB 060218 at 3 days. 
In terms of available prompt optical and afterglow radio flux, GRB 060218 seems to represent a common llGRB.



\section{Constraints on the prompt phase}\label{Sec:Prompt}
\subsection{Methodology}
\label{Sec:Methodology}
\subsubsection{Conditions on the comoving magnetic field}
\label{Sec:MagneticField}
For particles to successfully reach a specific energy, the acceleration time scale for that energy needs to be shorter than relevant cooling time scales \citep[e.g.,][]{Waxman1995, Murase2008, Guepin&Kotera2017}. The acceleration time scale for diffusive shock acceleration in a strong shock is of the same order as the Larmor gyration time and
is given by  $t'_{\textrm{acc},j} = \frac{E}{\eta c Z_j e B' \Gamma}$, where $E$ is the observed UHECR energy, $\eta$ is the acceleration efficiency\footnote{Often in the literature, the reciprocal of $\eta$ is defined as $\eta$. With our convention, larger $\eta$ equals faster acceleration.}, $Z_j$ is the charge number of particle species $j$, $e$ is the elementary charge, and $B'$ is the comoving magnetic field strength. In this paper, we only consider completely stripped iron with $Z = 26$, as we want to display the least constraining results. Accelerating protons and other lighter nuclei is more demanding. All primed parameters are evaluated in the comoving frame.

The particles cool mainly due to synchrotron radiation, the adiabatic expansion of the ejecta, and through interactions with the ambient photon field. The relevant time scales for these processes are $t'_{\textrm{sync},j} = \frac{6\pi}{Z_j^4 \sigma_{\textrm T}} \frac{(m_j c^2)^2}{c} \left ( \frac{m_j}{m_e} \right )^2 \frac{1}{ (E/\Gamma) {B'}^2}$, 
$t'_\textrm{ad} = \frac{r}{c\Gamma}$, and 
$t'_{p\gamma} = \frac{20 \pi r^2 \Gamma \left< \varepsilon \right>}{\sigma_{p\gamma}L_\gamma}$ respectively.\footnote{There can be a factor of a few difference in the expression of $t'_{p\gamma}$ depending on the photon spectrum at the source \citep[e.g.,][]{Murase2008}, but this does not influence our conclusion.}
In the above formulae, $\sigma_\textrm{T}$ is the Thomson cross section, $m_j$ is the mass of particle species $j$, $m_e$ is the electron mass, $r$ is the radial distance from the progenitor, $\left< \varepsilon \right>$ is a typical observed photon energy, and $\sigma_{p\gamma}$ is the photohadronic cross section. 
There are other ways in which the particles can lose energy but these three usually dominate and are sufficient for our present analysis \citep{Murase2006, Guepin&Kotera2017, Samuelsson2019}. Furthermore, considering additional cooling processes can only reduce the maximum UHECR energy.
The requirement $t'_{\textrm{acc},j} < \min [t'_{\textrm{sync},j}, \, t'_\textrm{ad}, \, t'_{p\gamma}]$ can be translated into constraints on the comoving magnetic field $B'$.\footnote{Additionally, another constraint can be put on the magnetic field as the magnetic energy output per time cannot be larger than the total luminosity \citep{Samuelsson2019}. As it turns out, this does not affect the conclusions presented in this paper and it is therefore not included in the current study. However, this should generally be considered when investigating UHECR sources.} 

\begin{figure}
\begin{centering}
    \includegraphics[width=0.9\columnwidth]{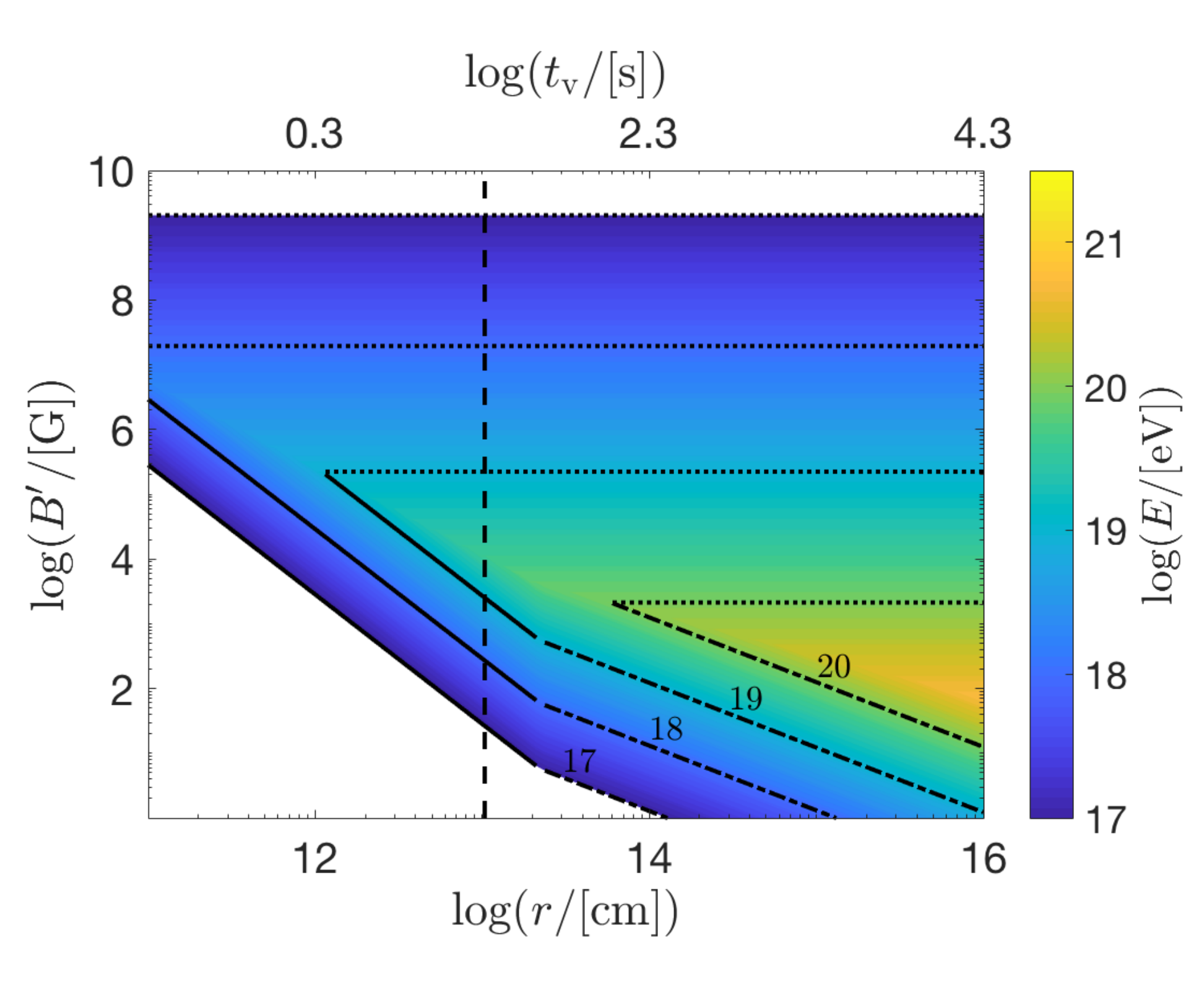}\hfill
    \includegraphics[width=0.9\columnwidth]{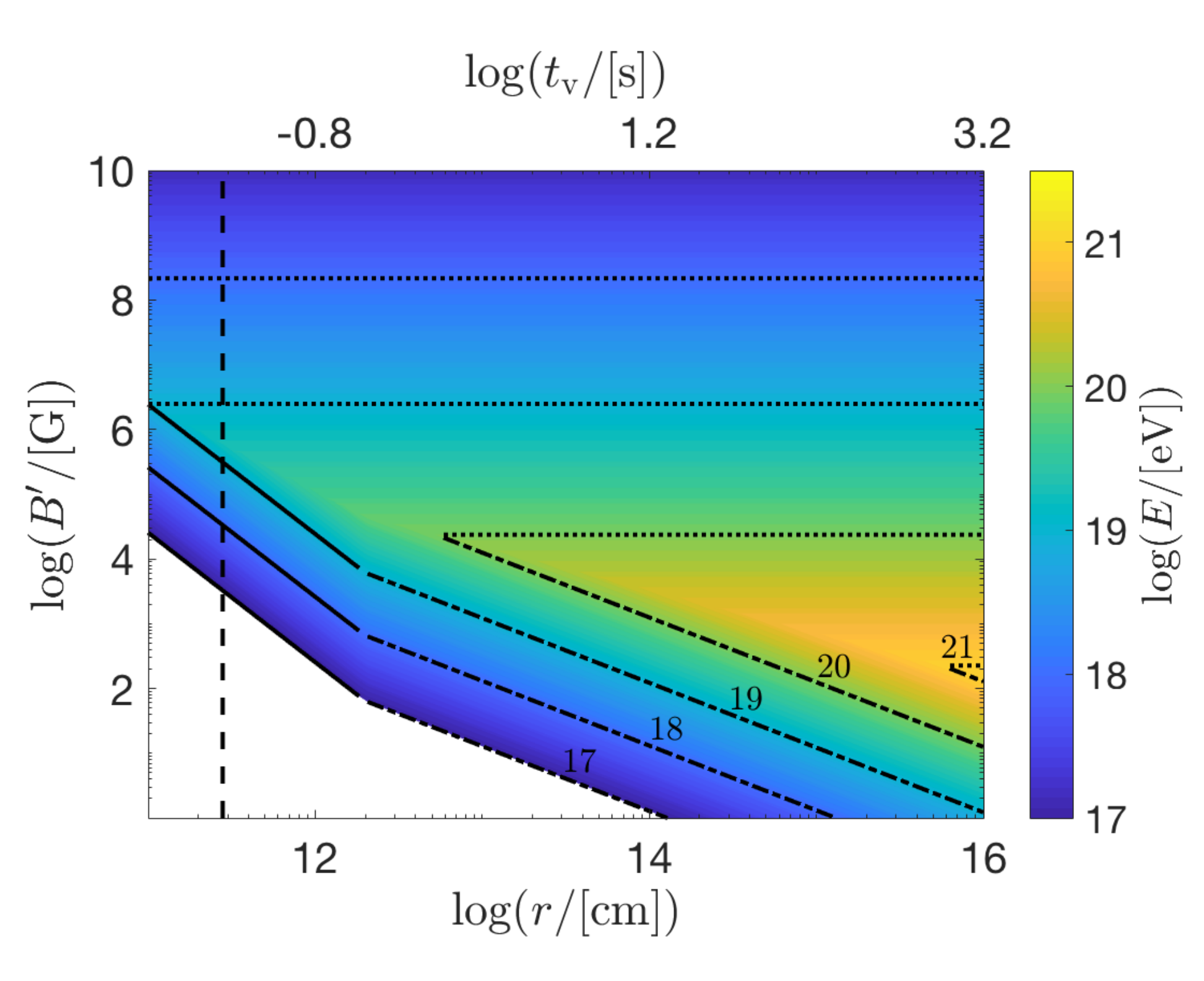}\hfill
    \includegraphics[width=0.9\columnwidth]{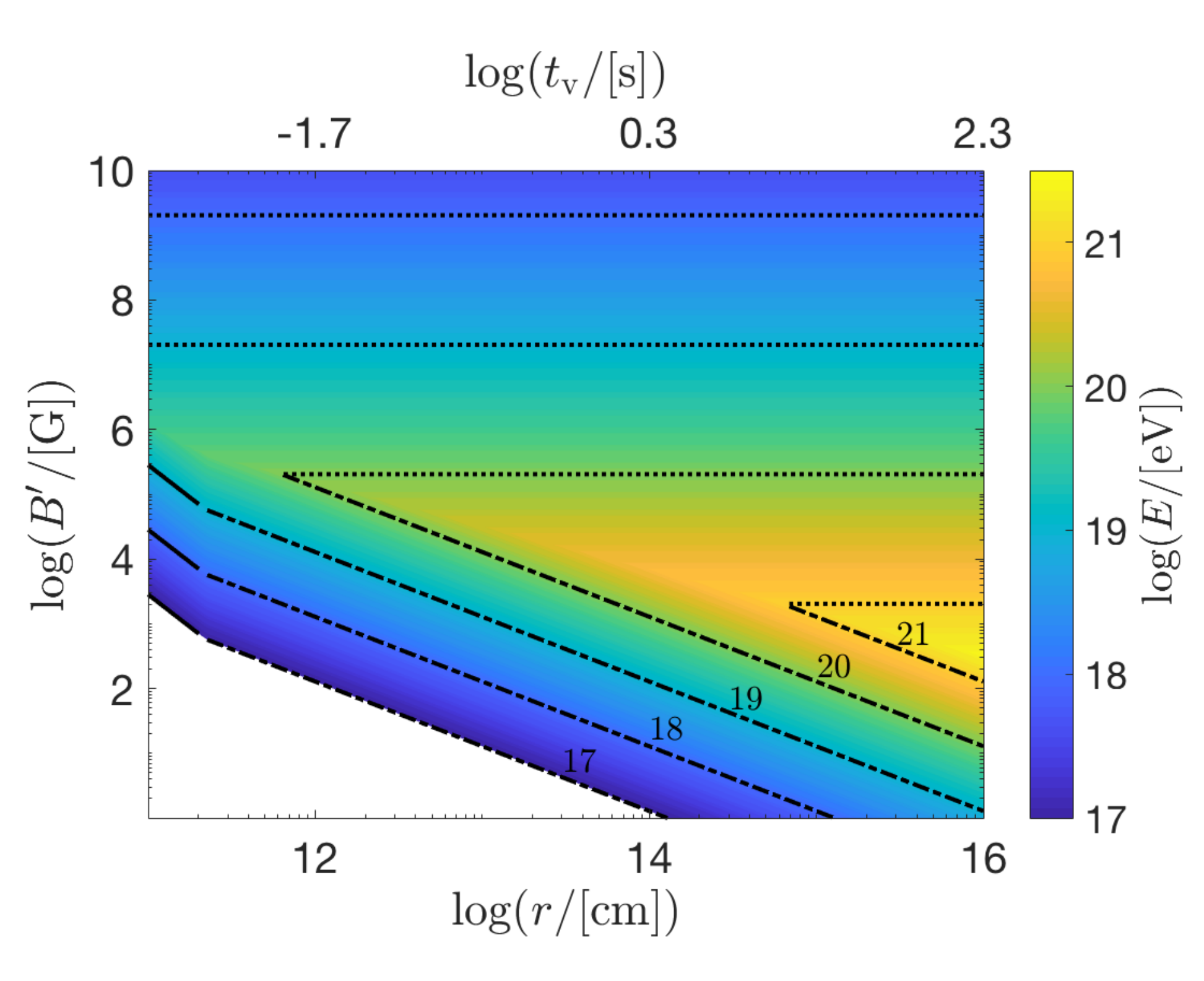}\hfill
\end{centering}
    \caption{Allowed parameter space for $B'$ as a function of $r$ for bulk Lorentz factor $\Gamma = 3$, 10, and 30 from top to bottom. Color bar shows $\log(E/[\textrm{eV}])$ for iron. Vertical dashed line shows the photosphere. Top $x$-axis shows the minimum variability time $t_{\textrm{v}} = r/2\Gamma^2c$. Dotted lines show the synchrotron limit, dotted-dashed lines show the adiabatic limit, and solid lines show the photohadronic limit, all for integer values of $\log(E/[\textrm{eV}])$ as indicated in the plots. 
Numerical values used are given in Table \ref{Tab:NumericalValues}.} 
    \label{Fig:MagneticField}
\end{figure}

The resulting parameter space for the magnetic field as a function of radius $r$ and maximum observed UHECR energy $E$ can be seen in Figure \ref{Fig:MagneticField}, plotted for 
$\Gamma = 3$, 10, and 30 from top to bottom. 
The $x$-axis extends up to $r = 10^{16}$ cm, comparable to the radio emission radius at 5 days \citep[$r \sim 3 \times 10^{16}$ cm,][]{Soderberg2006Nature}. 
The dashed vertical line shows the photosphere $r_{\rm ph} = L_\textrm{tot}\sigma_\textrm{T}/(8\pi m_p c^3 \Gamma^3)$ \citep{Peer2015}, where $m_p$ is the proton mass. Below the photosphere, particles are unlikely to be effectively accelerated \citep{LevinsonBromberg2008, Budnik2010, Murase2013, Beloborodov2017}. The top axis in Figure \ref{Fig:MagneticField} shows the minimum variability time $t_{\textrm{v}} = r/2\Gamma^2c$.

From Figure \ref{Fig:MagneticField} it is evident that 
there exists quite a large parameter space where UHECR acceleration would be possible. This is partly why GRBs and llGRBs have been extensively studied as promising UHECR candidates. 
For the figure, an acceleration efficiency of $\eta = 0.1$ has been used. 
This is well motivated by both theory and simulations \citep{Protheroe2004, Rieger2007, Caprioli2014}. However, the true value is not known and we explore the results for larger $\eta$ in Section \ref{Sec:Optimistic}. 

\subsubsection{Estimated synchrotron flux}
\label{Sec:SubFlux}
We will presently give a short description of how the synchrotron flux from the electrons in the UHECR acceleration region 
is estimated; a detailed explanation is given in \citet{Samuelsson2019}. As in  \citet{Samuelsson2019}, we work within the framework where electrons are instantaneously injected into a power law distribution as described by e.g., \citet{BlumenthalGould1970, Sari1998}. The effects of different mechanisms that could alter the electron distribution, such as stochastic acceleration, diffusion, or magnetic dissipation models, are outside the scope of this paper and left for future work. Synchrotron self-Compton (SSC) emission in the prompt spectra was investigated by \citet{Ghisellini2007}. They found that the XRT data, as well as the BAT data, do not allow for strong SSC emission. Therefore, we ignore the effects of SSC in this section. However, the requirement of weak SSC emission can be used to further constrain the viable parameter space for UHECR acceleration, and we discuss this possibility in Appendix \ref{App:InverseCompton}.

The characteristic energy of an emitted synchrotron photon depends on the Lorentz factor of the emitter. The observed synchrotron spectrum therefore depends on the electron distribution at the source. We assume that a number fraction $\xi_a$ of the electrons are injected into a power law with slope $-p$ between $\gamma'_\textrm{m}$ and $\gamma'_\textrm{max}$, where $\gamma'_\textrm{m} \ll \gamma'_\textrm{max}$. The minimum Lorentz factor is calculated as $\gamma'_\textrm{m} = a (m_p/m_e)(\epsilon_e^\textrm{NT}/\xi_a)$, where $a$ is a prefactor of order unity that we set to 1 and $\epsilon_e^\textrm{NT}$ is the fraction of internal energy given to nonthermal (NT) electrons.\footnote{We add the superscript NT to $\epsilon_e^\textrm{NT}$, as we use $\epsilon_e$ to describe the energy given to \textit{all} electrons in Section \ref{Sec:Afterglow}.} The conclusions are insensitive to if $a$ is increased, as appropriate in the case of e.g., magnetic reconnection, as most constraints are based on the cooler electrons (note that due to the high magnetic fields, we typically have the fast-cooling case leading to a soft-energy spectrum). 
The comoving number density of emitting electrons is calculated as \citep{Peer2015} $n_e' = \xi_a L_\textrm{tot}/(4\pi r^2 m_p c^3 \Gamma ^2)$. This value can be reduced by a large relative Lorentz factor among internal shocks or if dissipation lasts shorter than the dynamical time. A large relative Lorentz factor would only change the estimate by a factor of a few and not affect the conclusion much. For a short dissipation time, we again expect the conclusions not to change, as long as the low-energy electrons keep radiating over a dynamical time passively.

Electrons cool mainly through synchrotron emission or adiabatic cooling. This leads to a cooling break at $\gamma'_\textrm{c}$, defined as the electron Lorentz factor for which the energy losses from these two cooling processes are equal. Electrons can furthermore reabsorb synchrotron photons (synchrotron self-absorption), effectively heating any electrons that are below the characteristic electron Lorentz factor $\gamma'_\textrm{SSA}$. 
The shape of the electron distribution is set by the relative position of $\gamma'_\textrm{m}$, $\gamma'_\textrm{c}$, and $\gamma'_\textrm{SSA}$, which in turn determines the shape of the emission spectrum \citep{BlumenthalGould1970, Sari1998}. 

The observed break frequencies $\nu_\textrm{m}$, $\nu_\textrm{c}$, and $\nu_\textrm{SSA}$ typically emitted by electrons with Lorentz factors $\gamma'_\textrm{m}$, $\gamma'_\textrm{c}$, and $\gamma'_\textrm{SSA}$ respectively are \citep[see][]{Samuelsson2019}.
\begin{equation}\label{eq:Frequencies}
\begin{split}
    \nu_\textrm{m} &= 1.9\times 10^{19} ~\textrm{Hz} \ \Gamma_1 ~ B'_3 ~ (\epsilon_{e,-1}^\textrm{NT})^2 ~ \xi_{a,-2}^{-2}, \\[2.5mm]
    \nu_\textrm{c} &= 3.0\times 10^{11} ~\textrm{Hz} \ \Gamma_1^3 ~ r_{14}^{-2} ~ (B'_3)^{-3}, \\[2.5mm]
    \nu_\textrm{SSA} &= 1.8\times 10^{14} ~\textrm{Hz} \ L_{\textrm{tot},48}^{1/3} ~ \xi_{a,-2}^{1/3} ~ \Gamma_1^{2/3} ~ r_{14}^{-2/3}.
\end{split}
\end{equation}
The value and parameter dependencies of the synchrotron self-absorption frequency depend on the values of $\nu_\textrm{m}$ and $\nu_\textrm{c}$.

The normalization of the spectrum is set by the total emitted power per electron multiplied by the number of emitting electrons, scaled by the distance to the source as 
\begin{equation}\label{eq:F_max}
    F_\nu^\textrm{max} = 3.3~\textrm{Jy} \ L_{\textrm{tot},48} ~ \xi_{a,-2} ~ \Gamma_1^{-2} ~ r_{14} ~ B'_3,
\end{equation}
where a luminosity distance of $d_l=4.5\times10^{26}~$cm ($145~$Mpc) has been used. Once the shape and normalization of the spectrum are determined, one can calculate the predicted spectral flux $F_{\nu}^\textrm{theory}$ in any given electromagnetic band. For the parameters in Equation \eqref{eq:Frequencies}, electrons quickly cool from $\gamma'_\textrm{m}$ to $\gamma'_\textrm{SSA}$. Below  $\gamma'_\textrm{SSA}$ the electrons reabsorb photons, which stops them from cooling further. Thus, most electrons emit at the characteristic frequency $\nu_\textrm{SSA}$. Additionally, $\nu_\textrm{m}$ lies above the X-rays. The calculated spectral fluxes in optical-UV and X-rays are 
\begin{equation}\label{eq:Fluxes}
\begin{split}
    F_{\nu_\textrm{opt}}^\textrm{theory} &= F_\nu^\textrm{max} \left( \frac{\nu_\textrm{opt}}{\nu_\textrm{SSA}} \right)^{-1/2} \\
    &= 1.6~\textrm{Jy} \ L_{\textrm{tot},48}^{7/6} ~ \xi_{a,-2}^{7/6} ~ \Gamma_1^{-5/3} ~ r_{14}^{2/3} ~ B'_3, \\[2.5mm]
    F_{\nu_\textrm{X}}^\textrm{theory} &= F_\nu^\textrm{max} \left( \frac{\nu_\textrm{X}}{\nu_\textrm{SSA}} \right)^{-1/2} \\
    &= 40~\textrm{mJy} \ L_{\textrm{tot},48}^{7/6} ~ \xi_{a,-2}^{7/6} ~ \Gamma_1^{-5/3} ~ r_{14}^{2/3} ~ B'_3.
\end{split}
\end{equation}
In the example above, the coaccelerated electrons overshoot the observed optical flux by a factor of $3\times10^3$ and the X-ray flux by a factor of 400. The parameter dependence of the optical and X-ray fluxes depend on the positions of $\nu_\textrm{opt}$ and $\nu_\textrm{X}$ relative to $\nu_\textrm{m}$, $\nu_\textrm{c}$, and $\nu_\textrm{SSA}$.

The emitted synchrotron flux is, among other parameters, a function of the magnetic field. The requirements on $B'$ to support UHECR acceleration shown in Figure \ref{Fig:MagneticField} can thus be translated into requirements on the observed synchrotron flux. As there exist measurements of the flux in both the optical-UV and the X-ray band, it is straightforward to check if the flux from the electrons in the UHECR acceleration region is compatible with these observations or not.

For large values of $r$, following the prescription of \citet{Samuelsson2019}, the number of radiating electrons is overestimated. The reason for this is that in this paper, the total energy $E_\textrm{tot}$ is fixed to be the minimum $10^{51}$ erg required for sufficient UHECR production. This is in contrast to \citet{Samuelsson2019}, where the total energy was not specified. Therefore, in this paper it is possible for the number of radiating electrons $N_e$ to surpass the total number of electrons in the burst $N_{e,\textrm{tot}}$: the necessary condition $N_e < \xi_a N_{e, \textrm{tot}} = \xi_a \times \frac{E_\textrm{tot}}{\Gamma m_p c^2}$ breaks down for large enough $r$. 
Above this radius, marked in forthcoming figures by a blue dashed line, our calculated flux will start to progressively deviate from the true flux. Assuming $E_\textrm{tot} \sim T_{90}L_\textrm{tot}$, this occurs when the variability time $t_\textrm{v}$ becomes comparable to the prompt duration time, $t_\textrm{v} \sim T_{90}/2$. 
Emission from these radii will inevitably extend into the onset of the afterglow, due to the angular dependence on the radiation arrival time. The method used to constrain UHECR acceleration in the afterglow phase presented in Section \ref{Sec:Afterglow}, will naturally be valid for this part of the parameter space as well. Hence, we limit the results presented in this section to apply when $t_\textrm{v} < T_{90}/2$ only, treating larger radii as part of the afterglow phase. 


\subsection{Results}
\label{Sec:Results}
\subsubsection{Fiducial parameters}\label{Sec:Fiducial}
\begin{figure*}
\begin{centering}
    \includegraphics[width=0.65\columnwidth]{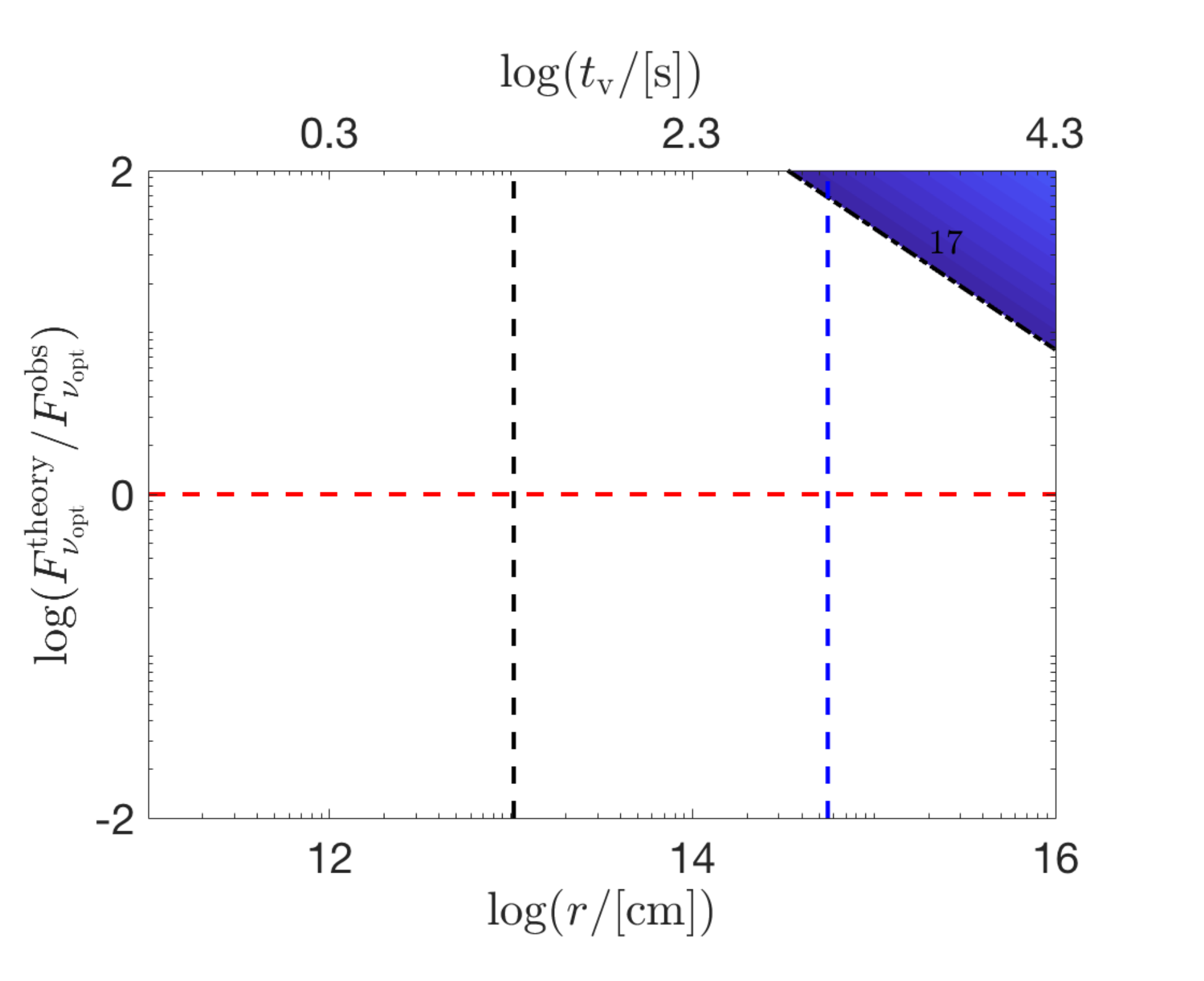}
    \includegraphics[width=0.65\columnwidth]{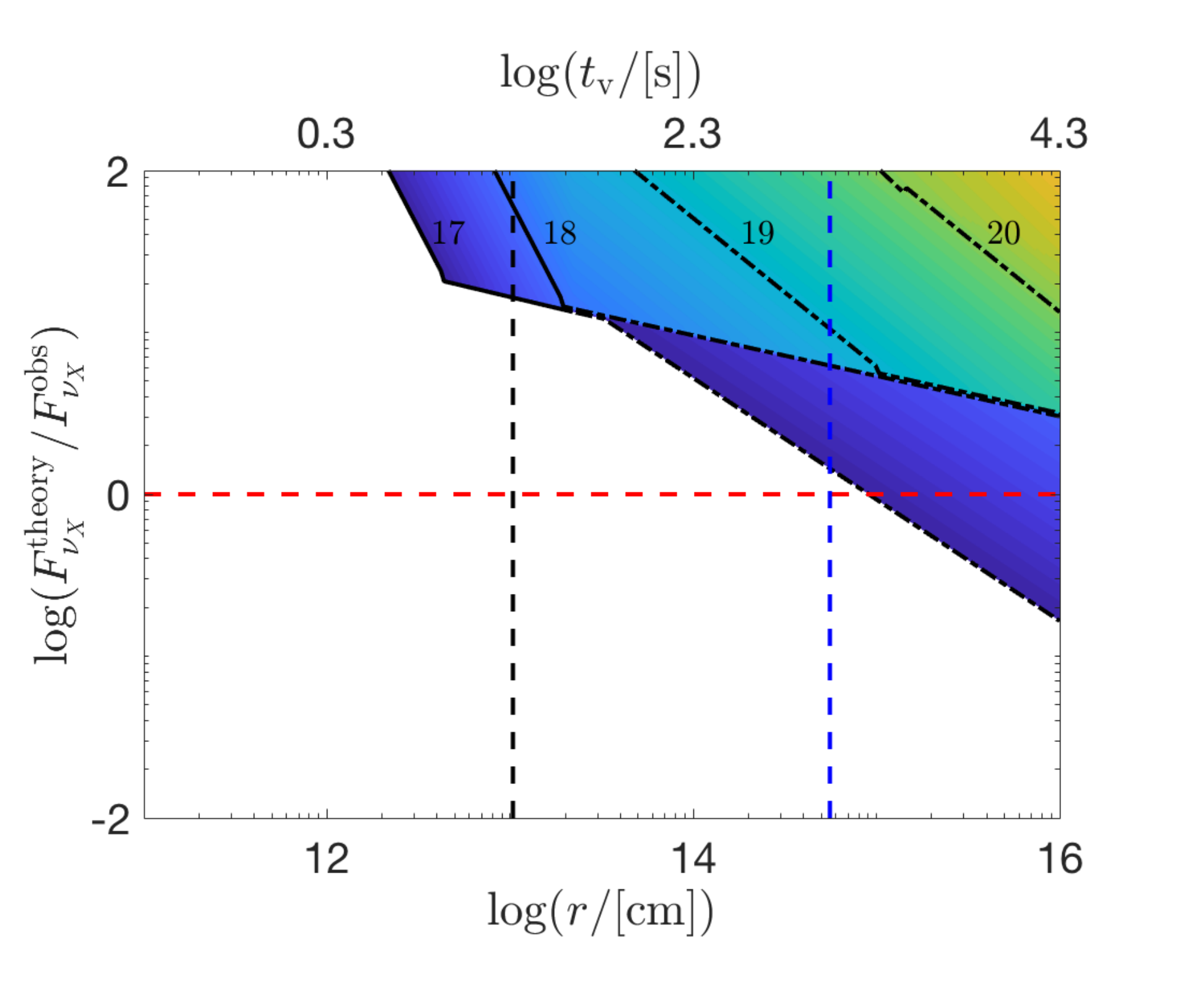}
    \includegraphics[width=0.65\columnwidth]{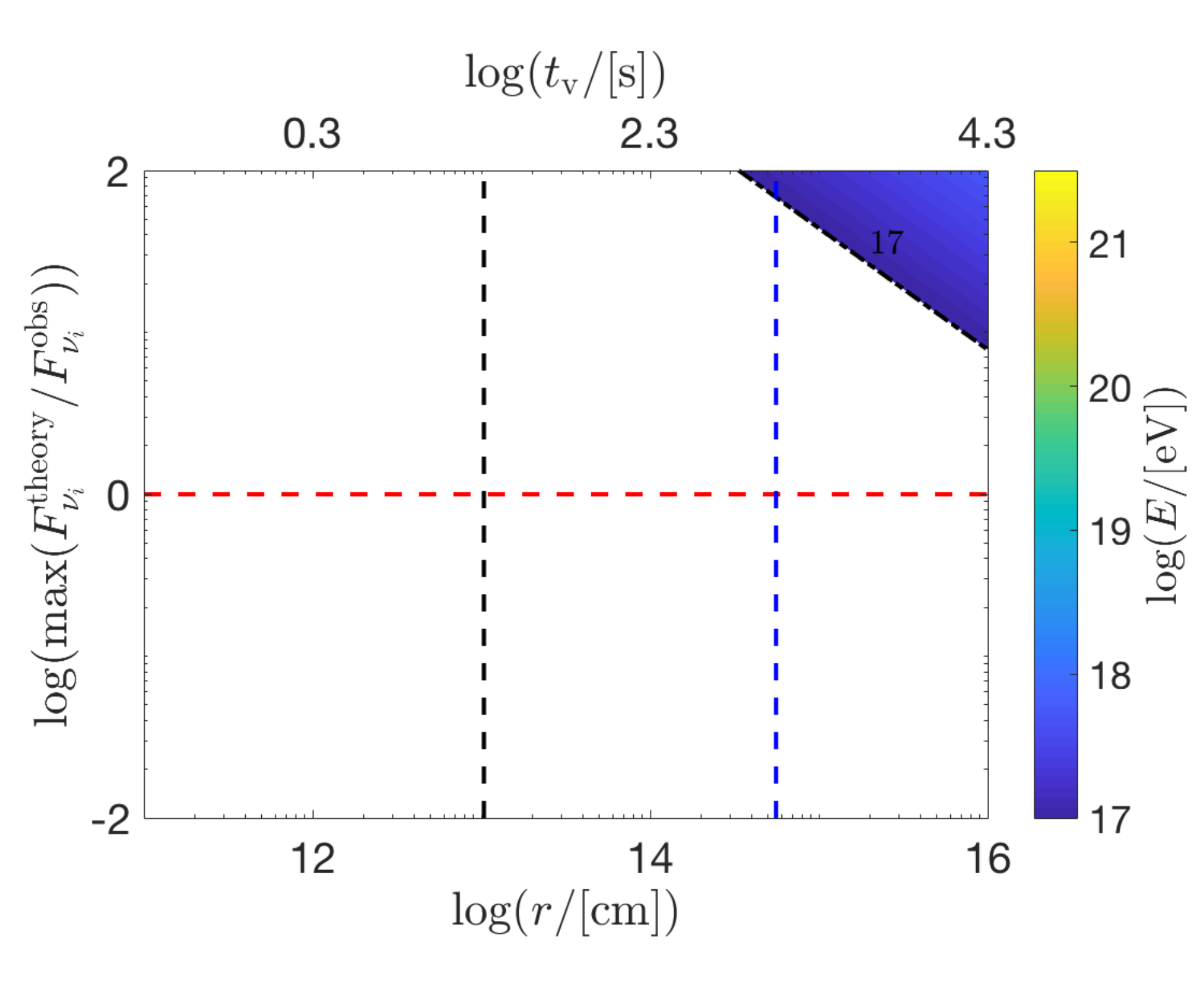}
    \includegraphics[width=0.65\columnwidth]{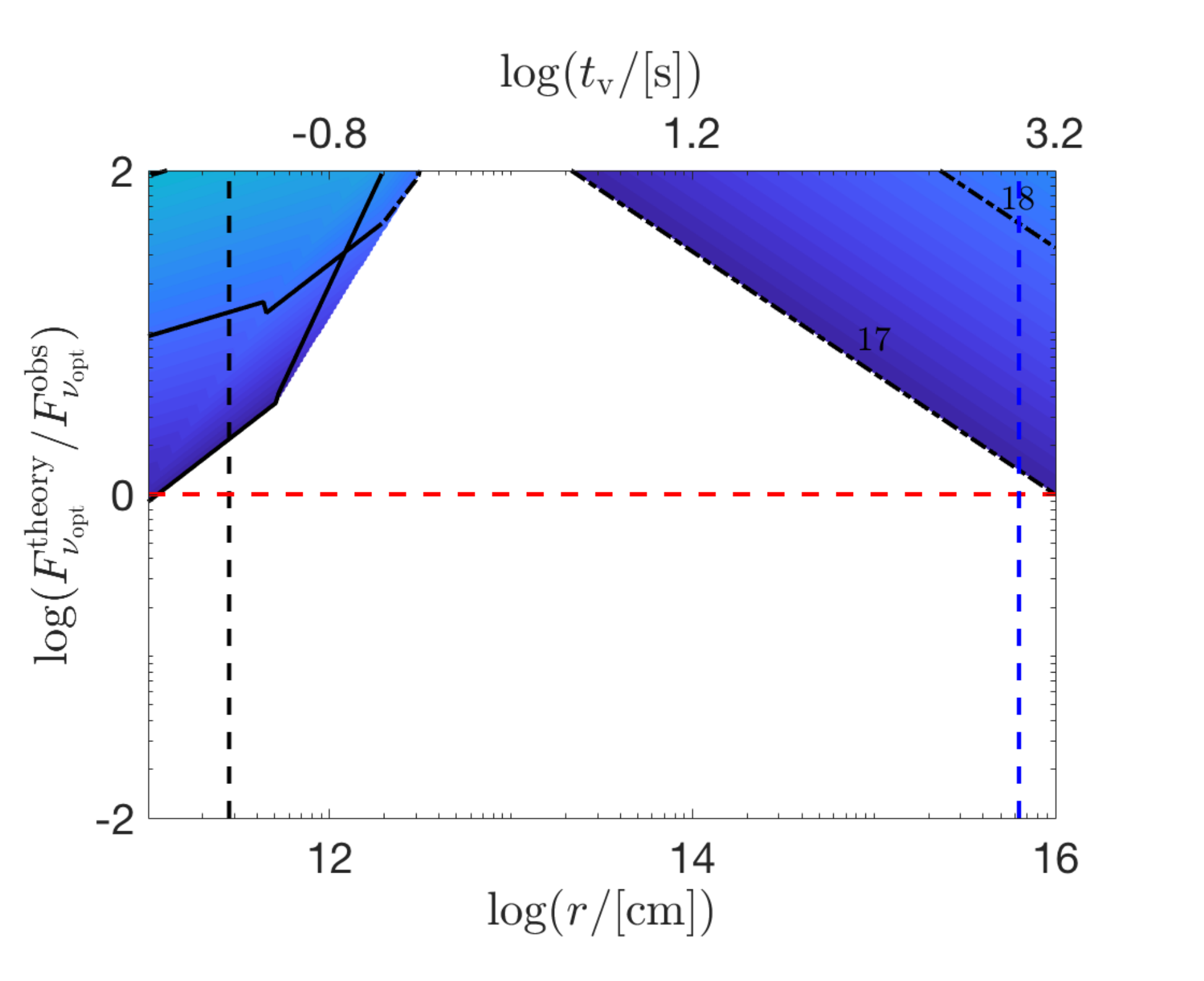}
    \includegraphics[width=0.65\columnwidth]{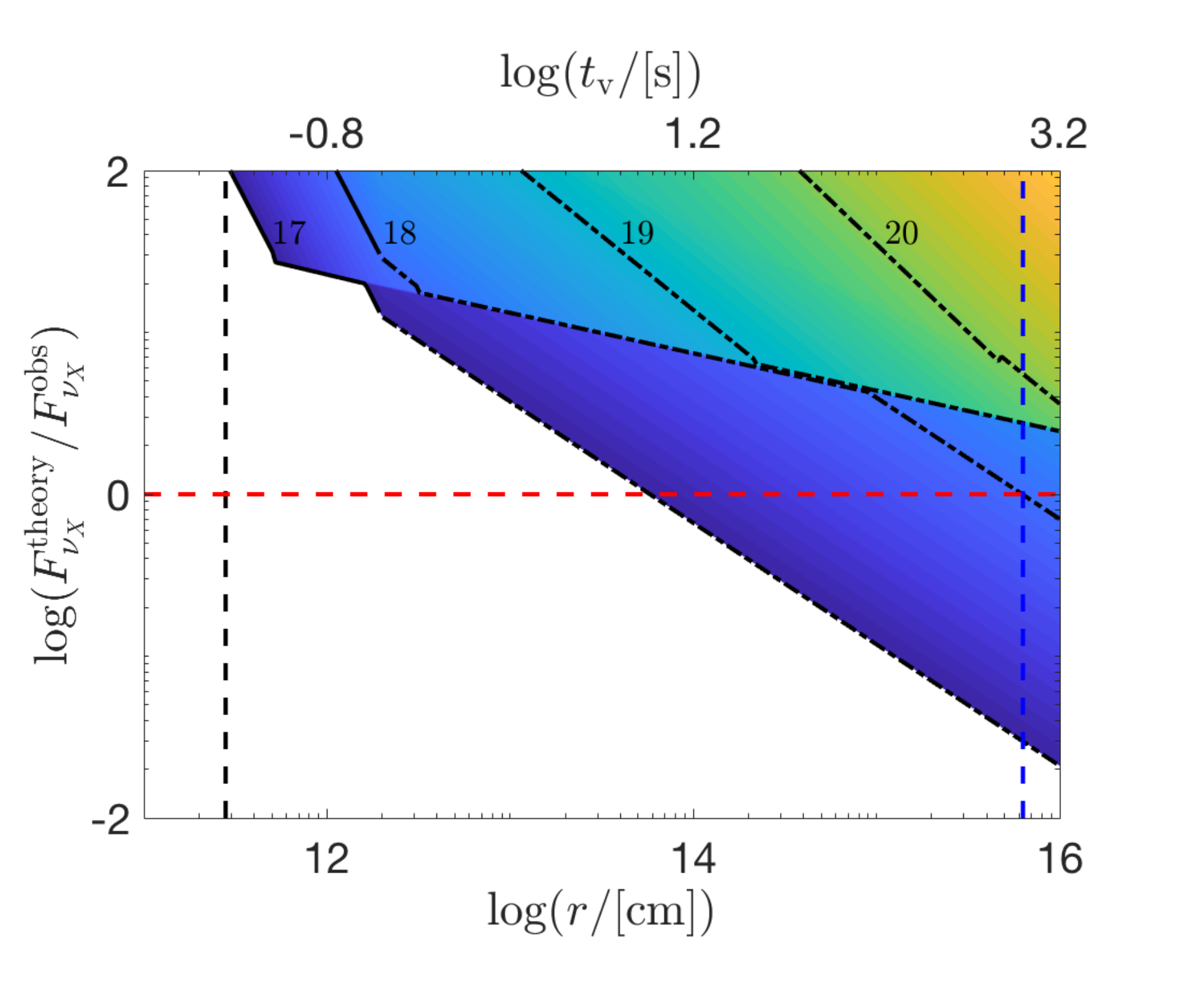}
    \includegraphics[width=0.65\columnwidth]{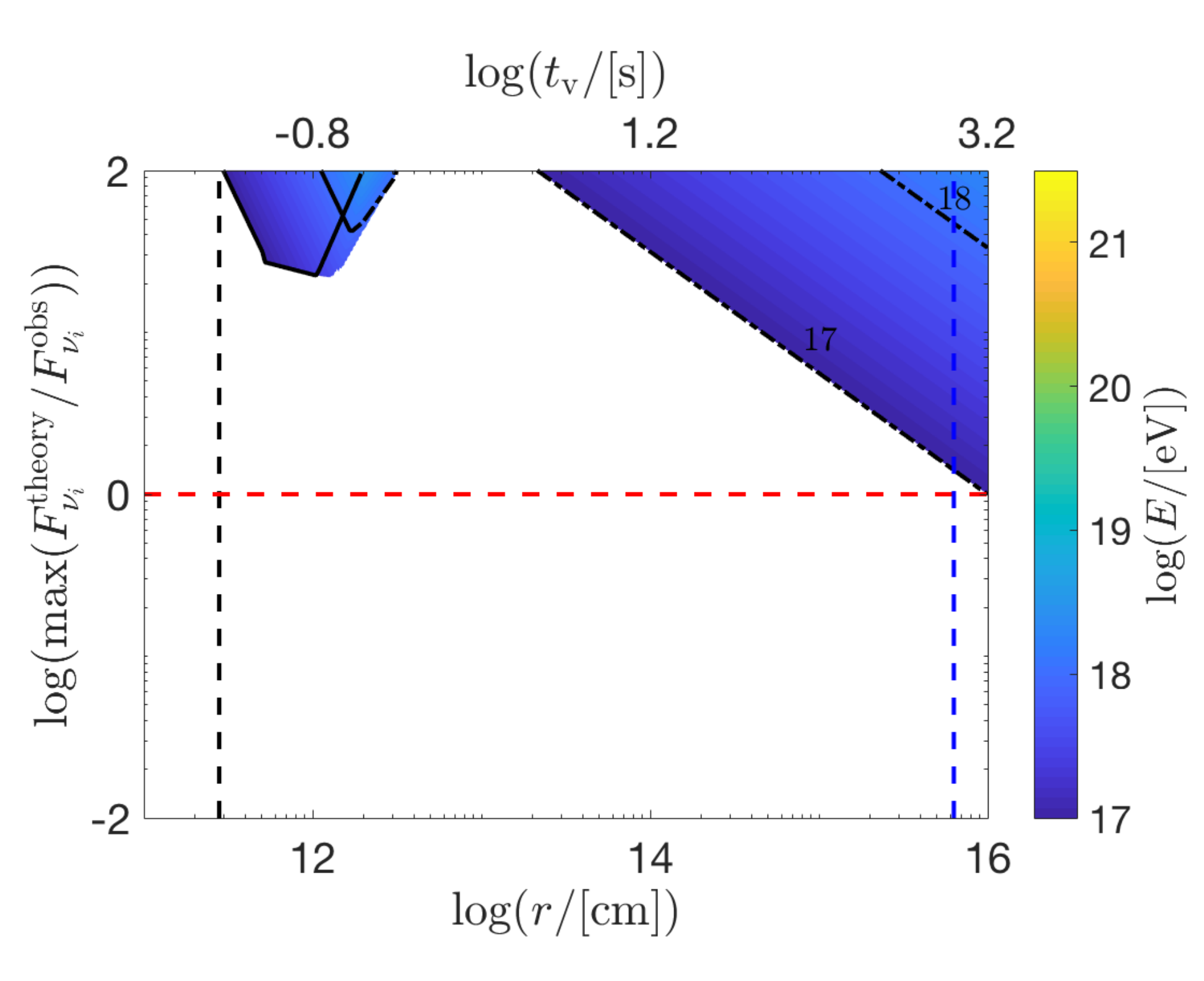}
    \includegraphics[width=0.65\columnwidth]{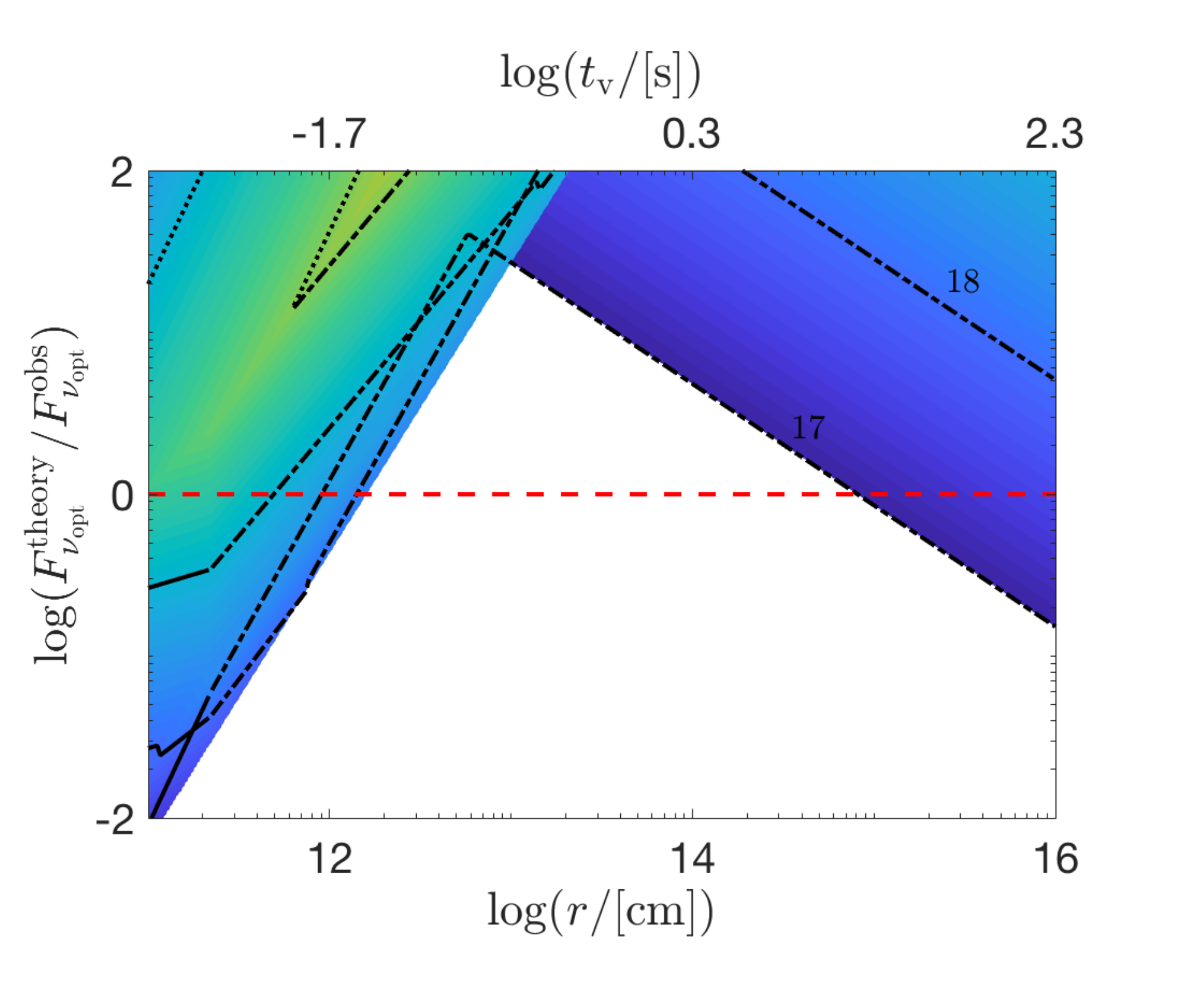}
    \includegraphics[width=0.65\columnwidth]{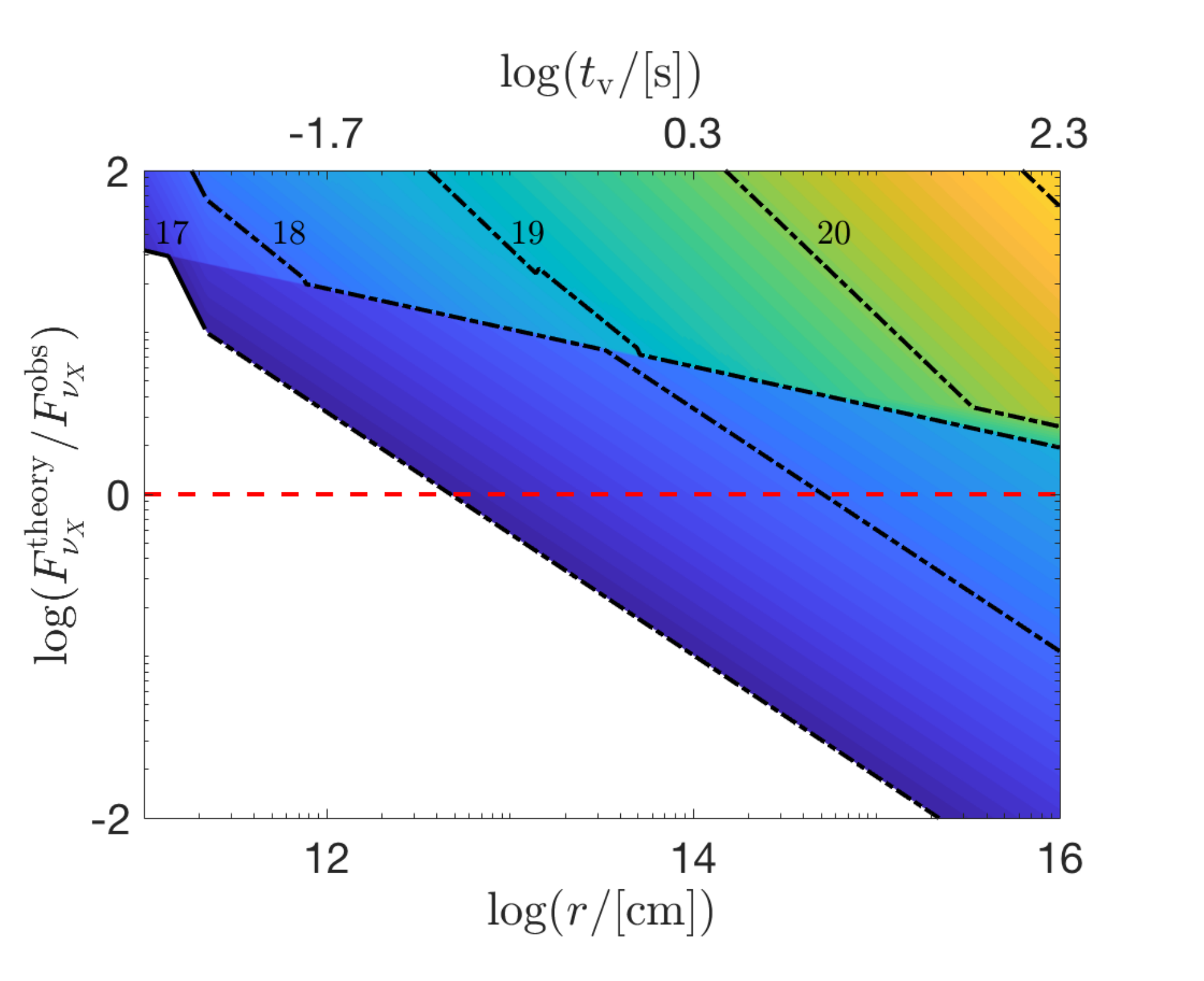}
    \includegraphics[width=0.65\columnwidth]{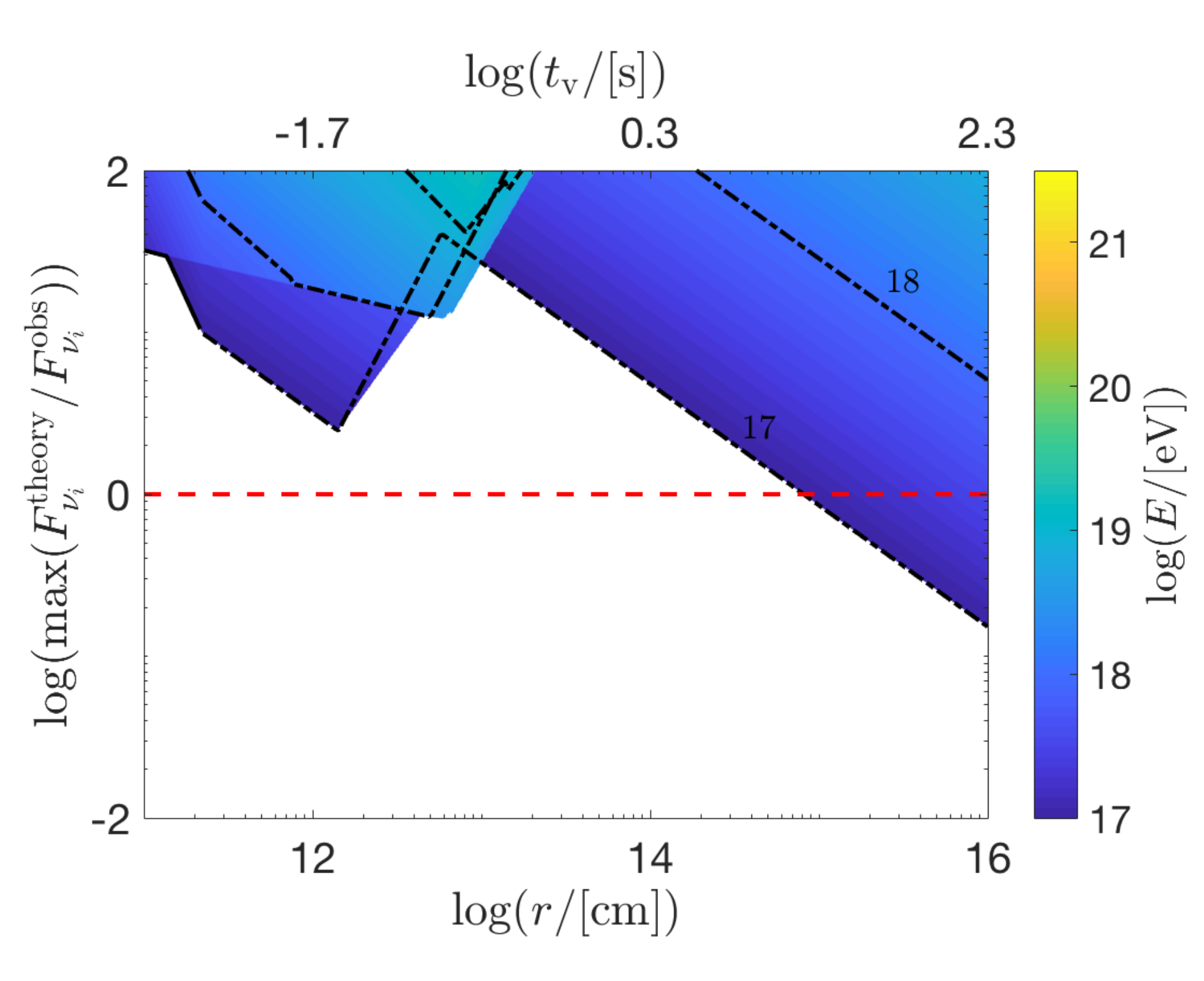}
\caption{Calculated synchrotron flux from the coaccelerated electrons in optical-UV (3 eV, left) and X-rays (5 keV, middle) normalized to their respective observational limit. On the right, both constraints are taken into account as $\max(F_{\nu_i}^\textrm{theory}/F_{\nu_i}^\textrm{obs})$ as the electron synchrotron flux has to be consistent with both the optical-UV and X-ray data. Plots are shown for $\Gamma = 3$, 10, and 30 from top to bottom, as a function of $r$ and $E$. 
Close to the progenitor, the optical flux is partially absorbed due to synchrotron self-absorption, while the peak flux is high. Optical flux increases rapidly when the absorption frequency decreases toward the optical band. For larger $r$, both the optical and the peak flux decrease, mainly due to the smaller allowed values of $B'$.
Red dashed line shows observational values, so everything above this is ruled out in our analysis. This implies that all of the parameter space for UHECR acceleration is ruled out, as it results in fluxes several orders of magnitude higher than observed, especially in the optical-UV band. Vertical, dashed, blue line shows $t_\textrm{v} = T_{90}/2$, above which our model begins overestimating the flux. We do not base any of our conclusion on this part of the parameter space. 
Line coding, color map, and top $x$-axis are similar to Figure \ref{Fig:MagneticField}.
Numerical values used are given in Table \ref{Tab:NumericalValues}.} 
\label{Fig:FluxLimits}
\end{centering}
\end{figure*}

The results for our fiducial parameters (given in Table \ref{Tab:NumericalValues}) are shown in Figure \ref{Fig:FluxLimits}. The synchrotron flux from the coaccelerated electrons in optical (left) and at 5 keV (middle) are both shown normalized to their respective observed deabsorbed flux (red dashed line). 
In the plots on the right, both constraints are taken into account as $\max(F_{\nu_i}^\textrm{theory}/F_{\nu_i}^\textrm{obs})$; the electron synchrotron flux has to be consistent with both the optical-UV and X-ray data. All of the parameter space above the red dashed line result in too bright synchrotron emission as compared to the observations and is thus ruled out.
From the figure, it is clear that no UHECR acceleration to energies $\gtrsim 10^{17}$ eV is possible. Specifically, the emission overshoots the observations in the optical band by several orders of magnitude, unless the emission radius is small. The vertical, dashed blue line shows  $t_\textrm{v} = T_{90}/2$, above which our model starts to overestimate the flux (see end of Section \ref{Sec:SubFlux}). 

\citet{Ghisellini2007} showed that the optical data can be explained as the self-absorbed part of a synchrotron spectrum. This explanation requires emission at small radii with $r \sim 10^{12}~$cm \citep[comparable to the radius expected from the shock breakout scenario][]{Waxman2007}. UHECR acceleration is prohibited at such small radii because of strong losses due to the photomeson production (\citealp{Murase2008}; see also Figure \ref{Fig:MagneticField}). However, efficient neutrino emission is expected, which could significantly contribute to the diffuse neutrino flux discovered by IceCube \citep[e.g.,][]{Murase2016, LevinsonNakar2020}.
\subsubsection{Optimistic parameters}\label{Sec:Optimistic}
Here, we investigate the set of parameters most favorable for CR acceleration. This effectively corresponds to the most conservative constraints on the prompt phase of GRB 060218-like transients as UHECR accelerators. The parameters that influence the results the most are the acceleration efficiency $\eta$, the number fraction of accelerated electrons $\xi_a$, and the fraction of internal energy given to the NT electrons $\epsilon_e^\textrm{NT}$ \citep[see][for details]{Samuelsson2019}. 
As previously mentioned, $\eta \sim 0.1$ is well motivated by both theory and simulations \citep{Protheroe2004, Rieger2007, Caprioli2014} but the true value is unknown. Increasing $\eta$ decreases the UHECR acceleration time, effectively increasing the allowed parameter space. However, if $\eta > 1$, the requirement that the Larmor radius of the particle should be smaller than the system size becomes constraining and must be taken into account \citep{Hillas1984, Waxman1995}, the effect of which is that higher UHECR acceleration will not be achieved by increasing $\eta$ beyond 1. 

For our fiducial parameters, we adopted the number fraction of accelerated electrons $\xi_a = 1$ as this is most commonly used \citep{Sari1998,Eichler2005,Santana2014}. However, the proper value of $\xi_a$ is likely lower.\footnote{Indeed, for slow shocks $\xi_a$ has to be small. This is because $\gamma'_\textrm{m} = a (m_p/m_e)(\epsilon_e^\textrm{NT}/\xi_a) \beta^2/2$, where $a$ is a numerical factor of order unity and $\beta$ is the shock velocity in units of the speed of light.
For nonrelativistic or trans-relativistic shocks, $\beta < 1$ and simulations suggest $\epsilon_e^\textrm{NT} \ll 1$. Hence, $\xi_a<1$ is required to keep $\gamma'_\textrm{m} > 1$.}
Decreasing $\xi_a$ has two effects in our analysis. Firstly, the maximum flux is proportional to the number of radiators so decreasing $\xi_a$ decreases the flux. Secondly, when fewer electrons are accelerated, they all receive a larger portion of the available energy, which increases $\gamma'_\textrm{m}$. Larger $\gamma'_\textrm{m}$ commonly leads to higher fluxes, especially in the higher energy bands. Therefore, it is not always true that a decrease in $\xi_a$ leads to less constraining results. 

The increase in $\gamma'_\textrm{m}$ can be counteracted by decreasing $\epsilon_e^\textrm{NT}$, as this decreases the energy available for the NT electrons. 
For relativistic shocks, the fractional energy $\epsilon_e^\textrm{NT}$ is well motivated to be $\sim 0.1$ \citep{Wijers1999,Panaitescu2000,SironiSpitkovsky2011,Santana2014} but for mildly relativistic outflows, $\epsilon_e^\textrm{NT}$ might be as low as $5\times10^{-4}$ \citep{Crumley2019}. Thus, llGRBs could potentially be dark cosmic ray accelerators, in the sense that UHECRs receive much more energy than electrons, which we assume dominate the radiation. 
In principle, when $\epsilon_e^\textrm{NT}$ is very small, previously subdominant radiation processes such as proton synchrotron emission or emission from photohadronic cascades could become non-negligible. Thus, additional constraints might be obtained by considering these processes as well. However, this is out of the scope of the current paper. 

\begin{figure*}
\begin{centering}
    \includegraphics[width=0.65\columnwidth]{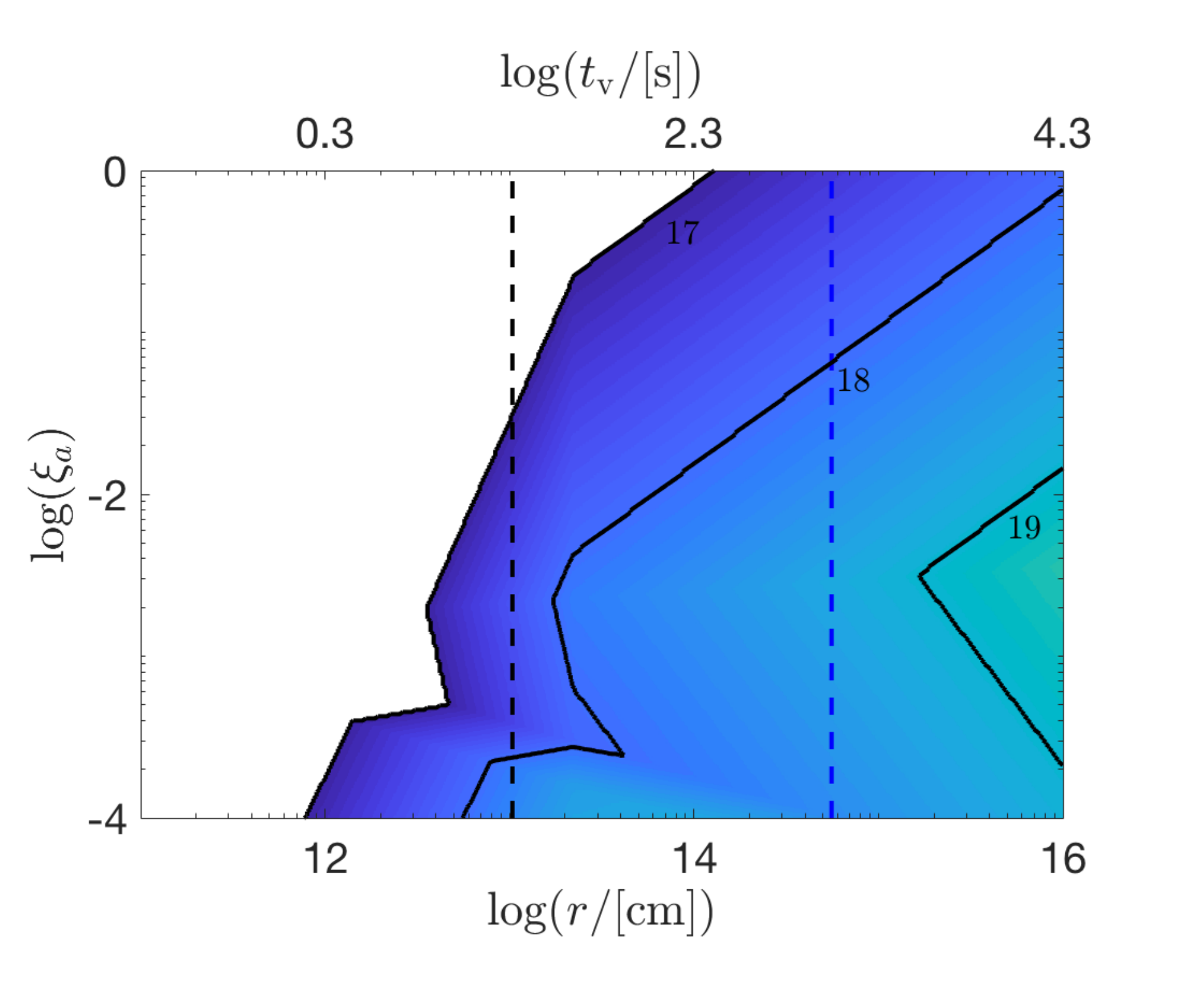}
    \includegraphics[width=0.65\columnwidth]{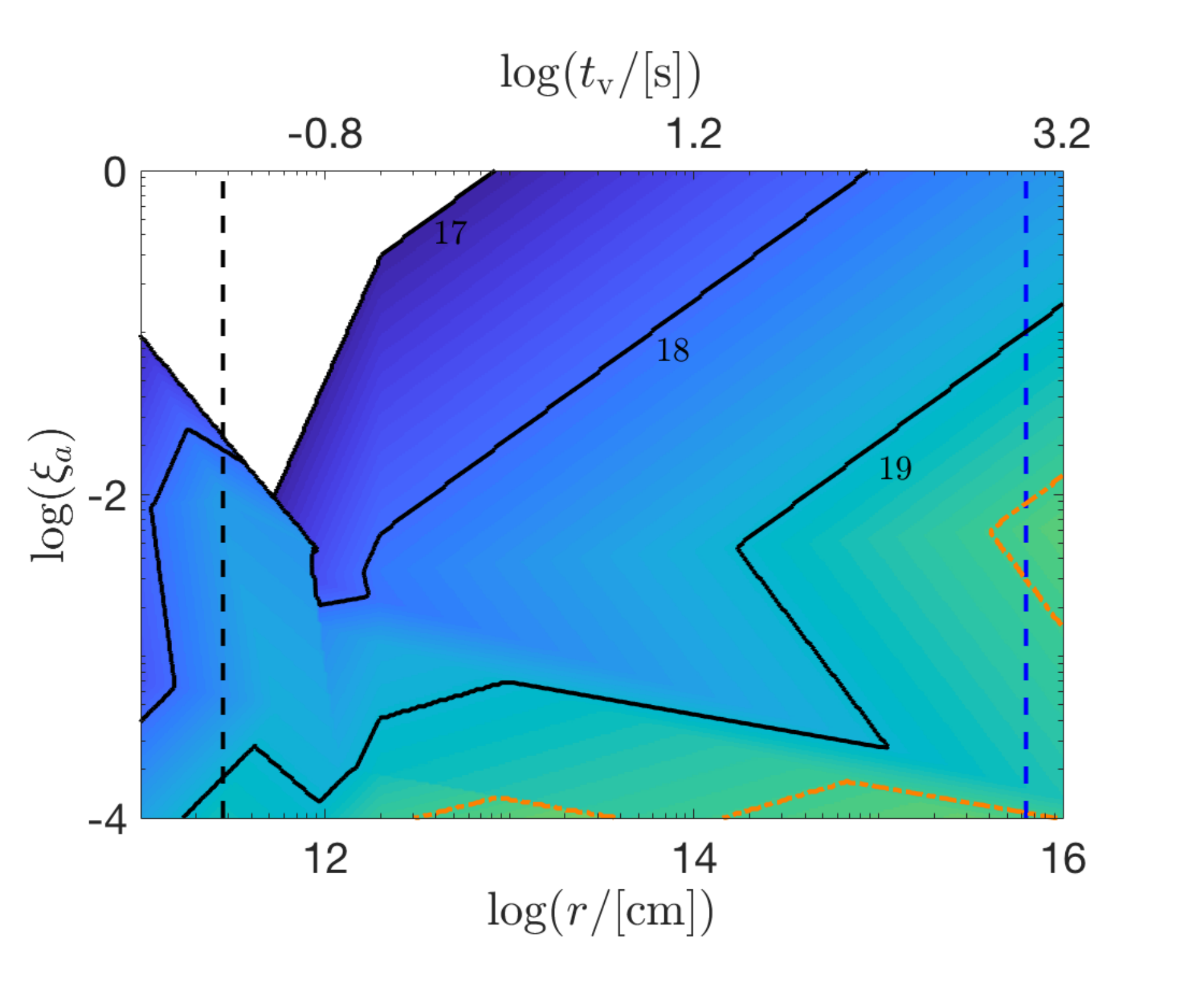}
    \includegraphics[width=0.65\columnwidth]{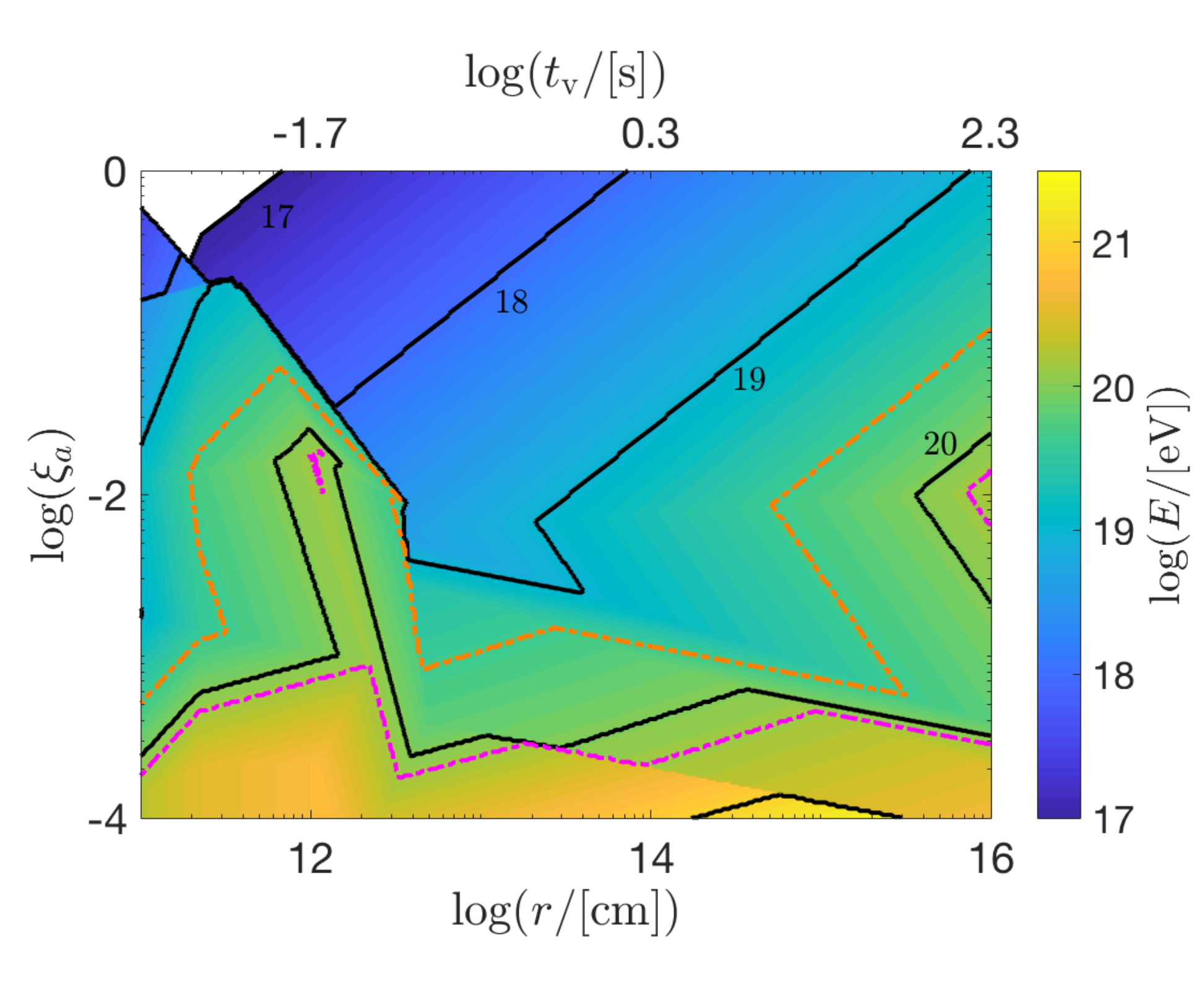}
    \caption{Maximum observed iron energy as a function of $\xi_a$ and $r$ for bulk Lorentz factor $\Gamma = 3$, 10, and 30 from left to right, for our optimistic parameters. The behavior is nontrivial, with synchrotron emission and photohadronic processes limiting the maximum energy for small $r$, while adiabatic cooling is the dominating limitation above $\sim 10^{13}$ cm. 
    Plots are made assuming high acceleration efficiency $\eta = 1$, and small fractional energy given to electrons $\epsilon_e^\textrm{NT} = 5\times 10^{-4}$. Orange and pink dotted-dashed lines show the maximum energy sufficient to fit the UHECR spectrum, as found by \citet{Heinze2019} and \citet{PierreAugerJCAP2017} respectively. Color map and top $x$-axis are similar to Figure \ref{Fig:MagneticField}. Previous line coding has been dropped, with the solid black lines now corresponding to the most constraining of the synchrotron, photohadronic, and adiabatic limits. Vertical, dashed blue line shows $t_\textrm{v} = T_{90}/2$, above which our model begins overestimating the flux. We do not base any of our conclusion on this part of the parameter space. Other numerical values used are given in Table \ref{Tab:NumericalValues}.}
    \label{Fig:Robustness}
\end{centering}
\end{figure*}

\begin{figure}
    \includegraphics[width=1\columnwidth]{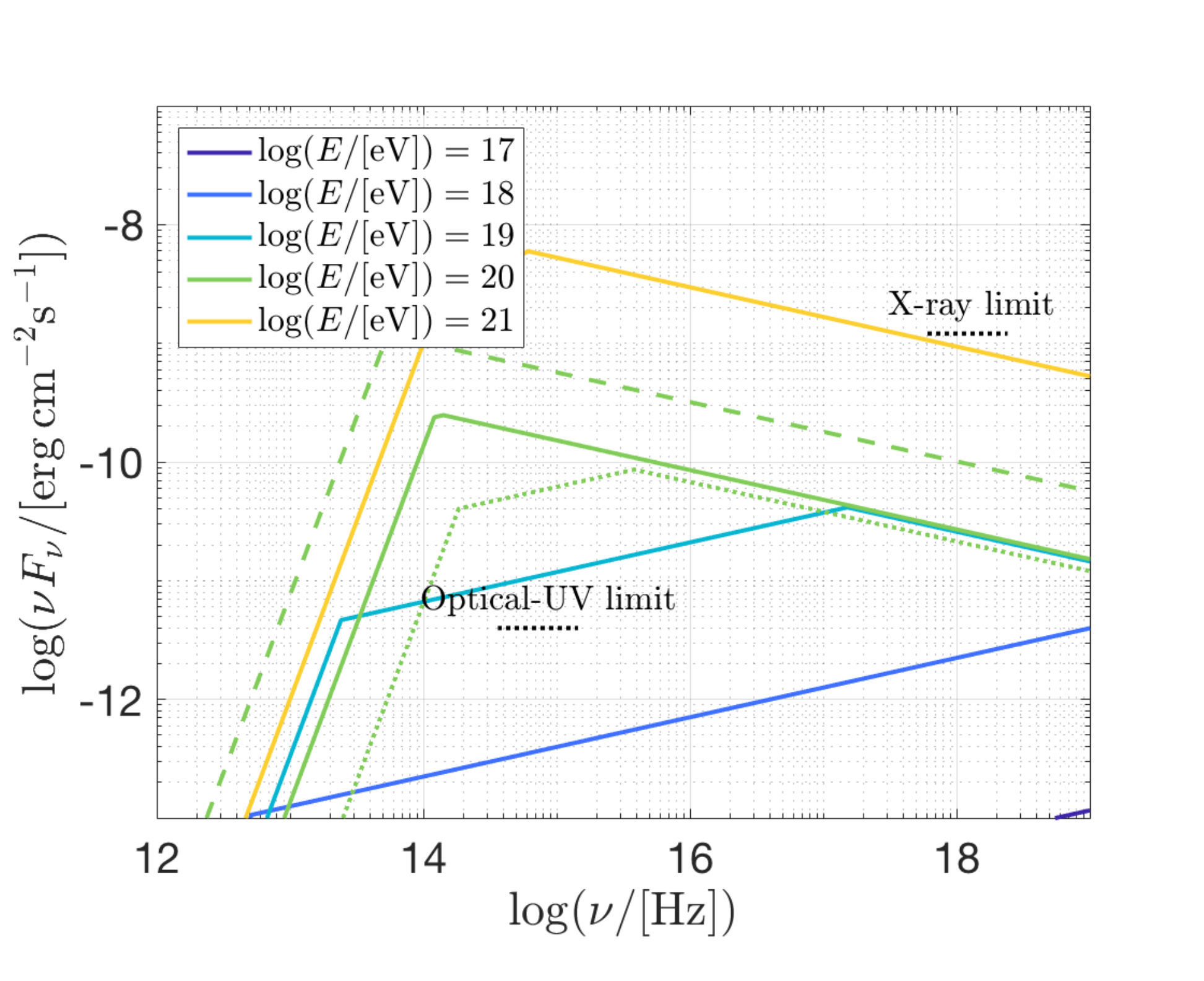}
    
    \caption{Spectra for integer values of $\log(E/[\textrm{eV}])$ using our optimistic parameters ($\eta = 1$ and $\epsilon_e^\textrm{NT}=5\times10^{-4}$). Spectra are calculated at $r = 10^{14}~$cm and for $\xi_a = 10^{-2}$. Solid lines are for $\Gamma = 10$. For visibility, we only show the $\log(E/[\textrm{eV}]) = 20$ spectrum for $\Gamma = 3$ (dashed) and $\Gamma = 30$ (dotted). Flux limits in optical-UV and X-rays used in this work are shown by black dotted lines \citep{Campana2006, Ghisellini2007}. For the highest cosmic ray energies, the spectra peak close to the optical band, making this observational band a good diagnostic of UHECR acceleration. See \citet{Samuelsson2019} for details of how the spectra are calculated. Other numerical values used are given in Table \ref{Tab:NumericalValues}.}
    \label{Fig:Spectra}
\end{figure}

For our optimistic parameters, we put $\eta = 1$, $\epsilon_e^\textrm{NT} = 5\times10^{-4}$, and let $\xi_a$ be a free parameter, to see what maximum iron energy is achievable while still being consistent with both the optical-UV and X-ray constraints. In Figure \ref{Fig:Robustness}, we show the result. Even in this case, with very high acceleration efficiency, completely stripped iron nuclei, and very little energy received by the NT electrons, decreasing the fraction of radiating electrons down to $\xi_a = 10^{-4}$ is not sufficient to reach the highest energies of $10^{20}$ eV when $\Gamma = 3$ and 10. The synchrotron emission from the electrons is still too bright, outshining the measurements in the optical band. For $\Gamma = 30$, UHECR acceleration to the highest energies is not constrained for small values of $\xi_a$. However, the result relies on the internal shocks remaining mildly relativistic for such large values of $\Gamma$, which is uncertain. 
In Figure \ref{Fig:Spectra}, we show examples of expected spectra for $r = 10^{14}~$cm and $\xi_a = 10^{-2}$, using our optimistic parameters. Details of how the spectra are calculated can be found in \citet{Samuelsson2019}. From the figure, it is clear that prompt optical-UV data is crucial to constrain the electron synchrotron spectrum and, in extension, the possibility of UHECR acceleration.

\subsection{Discussion}\label{Sec:DiscussionPrompt}

For $\Gamma = 30$ and $\xi_a \sim 10^{-2}$ or lower, we cannot constrain UHECR acceleration. However, when modeling of prompt GRB emission in the context of shocks one often assumes that $\epsilon_e^\textrm{NT} \sim 0.1$, such as to overcome the low radiative efficiency problem. Accepting this as a constraint together with the BAT data, the possible parameter space disappears. This is mostly because it is difficult to accommodate the low-energy slope of the spectrum (and so the optical data) with fast-cooling synchrotron spectrum that predicts a soft low-energy slope. The information from BAT is necessary because for small values of $\xi_a \lesssim 10^3$, the X-ray data can be compatible but the $\nu F_\nu$-peak would then fall above $100~$keV.

In a recent paper, \citet{Heinze2019} have shown that the maximum energy required to fit the UHECR spectrum can in fact be lower than $10^{20}$ eV. They get their best fit for a maximum rigidity of $R_\textrm{max} = E_\textrm{max}/eZ = (1.6\pm 0.2) \times 10^{18}$ V, which translates into a maximum energy of $4\times 10^{19}$ eV for stripped iron nuclei. Their work used a similar method to that of \citet{PierreAugerJCAP2017}, who found a higher best fit rigidity corresponding to an iron energy of $1.4 \times 10^{20}$ eV. The difference arises from the different propagation codes used, with underlying uncertainties regarding the nuclear physics and the extragalactic background light modeling. Furthermore does \citet{Heinze2019} allow for the UHECR source population to evolve with redshift as compared to \citep{PierreAugerJCAP2017}. 
To account for the uncertainty in the maximum energy required, we have marked these energies with dotted-dashed lines in Figure \ref{Fig:Robustness}. For $\Gamma = 10$, $4\times 10^{19}~$eV can be obtained in a fine-tuned region when $r > 5\times10^{15}$ cm and $\xi_a \lesssim 10^{-2}$, or for smaller values of $r$ if $\xi_a < 2\times 10^{-4}$.

Previous studies \citep{Murase2006,Murase2008, Liu2011, Zhang2018} have considered the observed photon spectrum to compute the cooling of UHECR through photohadronic interactions. As such, they have obtained constraints on the magnetic field, finding a parameter space compatible with the acceleration of UHECR. In those studies, the phenomenological shape of the photon spectrum was assumed. Specifically, a hard low-energy slope was used. Such a spectrum has been known to be incompatible with the simple fast-cooling spectrum. We stress that in this work, we do not aim to reproduce the observed spectrum, but simply require the predicted fluxes to be compatible with observations. 
The requirements on any dissipation model that could account for the production of UHECR and replicate the spectrum requires that a small fraction of the electrons be maintained in a power law with a steep electron index $p \lesssim 1$. Our result gives important general constraints on cosmic ray acceleration in llGRBs 
and we leave it for future work to inquire what constraint can be put on such models.

When $\xi_a$ is small, the bulk number of the electrons is thermal. In this section, we neglect the contribution of these electrons to the absorption and emission as it will not affect our conclusion of the results presented here. Note that we do include their contribution when studying the afterglow phase in Section \ref{Sec:Afterglow}. 
\citet{Warren2017, ResslerLaskar2017,Warren2018} have all studied the effects of the thermal electron population on the observed spectrum and found that they radiate mostly in the lower energy bands, leading to an increase of the flux in optical and radio. These studies were made in the context of afterglow emission, and thus might not be applicable to the prompt phase. However, they indicate that a more detailed treatment of the thermal population of electrons might result in stronger constraints from the optical-UV flux. 

Our model becomes inaccurate for $r$ satisfying $t_\textrm{v} \gtrsim T_{90}$, as mentioned in Section \ref{Sec:SubFlux}. Inspection of Figures \ref{Fig:FluxLimits} and \ref{Fig:Robustness} shows that this is of more concern for $\Gamma = 3$ than for $\Gamma = 10$ and 30. Incidentally, this is also when our constraints are strongest. Furthermore, the X-ray flux starts dropping rapidly after it peaks at 1000 s, having dropped by one (two) order(s) of magnitude after $\sim 5000$ s ($\sim 7000$ s) \citep{Campana2006}. If one wish to evaluate UHECR acceleration at larger radii using the methodology of this section, one would need to account for this variation in $F_{\nu_\textrm{X}}^\textrm{obs}$. This would lead to much stronger constraints from the X-ray flux once $t_\textrm{v} \sim 5000$ s, corresponding to $\sim3 \times 10^{15}$ cm for $\Gamma = 3$.

The prompt optical luminosity of GRB 060218 is higher compared to the other two available accounts for llGRBs in the literature \citep[llGRBs 100316D and 171205A,][]{Starling2011,Fan2011,DElia2018}. Because of the slew time of optical instruments, it is difficult to get measurements during the prompt phase. Thus, the sample of llGRBs with prompt optical detections will be biased toward long-duration bursts. Interestingly enough, GRB 171205A with an optical luminosity $\sim 6$ times smaller than GRB 060218, belongs to the category of shorter and harder llGRBs. This hints at a common distribution of prompt optical luminosities for all llGRBs. Furthermore, it seems that GRB 060218 might lie in the upper part of that distribution. If the prompt optical luminosities generally are lower, then our constraints on UHECR acceleration in the prompt phase of llGRBs would become even stronger. 


\section{Constraints on the afterglow phase}\label{Sec:Afterglow}
So far, we have only dealt with acceleration of UHECRs during the prompt phase. When the kinetic energy of the swept up material equals the rest-mass energy of the outflow, 
the blast wave starts to decelerate substantially. The interaction between the outflow and the circumburst medium creates a forward and a reverse shock, which convert the kinetic energy to internal energy. 


    

Previous modeling of the late-time emission of GRB 060218 as forward shock emission have required a rather small kinetic energy in the blast wave to be consistent with observations ($\sim 10^{48}\textrm{--}10^{50}$ erg) \citep{Soderberg2006Nature, Fan2006, Toma2007}. As we need at least $10^{51}$ erg to supply the UHECR flux, their solutions are not directly applicable. However, as pointed out in \citet{Eichler2005}, there may exist a degeneracy in the afterglow diagnostics. If the number of electrons accelerated into the power law is decreased, a GRB can be intrinsically more energetic without changing the observables. A small number fraction of accelerated electrons is known to be realistic in the case of nonrelativistic shock acceleration. As we need $\sim 10\textrm{--}100$ times more energy in the blast wave compared to the working solutions of \citet{Soderberg2006Nature, Fan2006, Toma2007}, we require the number fraction of accelerated electrons to be $\xi_a \sim 10^{-2} \textrm{--} 10^{-1}$ following the parameterization of \citet{Eichler2005}. 
The way to break the observable degeneracy is to study the emission from the bulk population of electrons ($1-\xi_a$) that we assume constitutes a thermal distribution in the downstream of the shock. 

A recent paper by \citet{ZhangMurase2019} studied the possibility of accelerating UHECRs at the reverse shock of GRB 060218-like transients. They consider two representative scenarios. In the first scenario (TRSN models), $>10^{51}$ erg is carried by a trans-relativistic component (with $\Gamma \beta<1$). This case is not constrained by the present work because UHECR production occurs at late times. 
In the second scenario, all UHECRs are assumed to be produced by a mildly relativistic jet ($\Gamma \sim 2\textrm{--}10$) component. In this case, they assume $\xi_a \sim  10^{-2} \textrm{--} 10^{-1}$, and here we show that strong constraints can be placed on these jet afterglow models. 

\citet{ZhangMurase2019} find that the observed UHECR spectrum and composition can be well reproduced considering their acceleration at the external reverse shock, and that the neutrino signal produced would not overshoot the diffuse flux detected by IceCube. Furthermore, they do calculate the secondary electromagnetic emission from the coaccelerated electrons 
and find that this is within observational constraints. 
However, there is a key, underlying assumption in \citet{Eichler2005}, which affects the electromagnetic signal calculated in \citet{ZhangMurase2019}. The assumption is that the nonaccelerated electrons are cold compared to the NT ones, which means that their synchrotron emission can be safely ignored. In reality, plasma waves are expected to transfer energy from protons to electrons, so that the thermal electrons are ``heated''. This notion is supported by PIC simulations, which show a continuous transition between the thermal and NT electrons \citep{SironiSpitkovsky2011, Park2015, Crumley2019, Bohdan2019}. Emission from the thermal electrons is not considered in \citet{ZhangMurase2019}, while we in this work demonstrate its importance.


\subsection{Methodology}\label{Sec:MethodologyAG}
To properly capture synchrotron emission and absorption from both the thermal and NT electron populations, we resort to numerical simulations. Available radio data from GRB 060218 suggests that there is a break in the spectrum at $\lesssim$ 10 GHz corresponding to the self-absorption frequency. This can be seen from the inset in Figure 1 of \citet{Soderberg2006Nature} as well as from the available data at 15.0 and 22.5 GHz at $\sim$ 3 days (also see Figure \ref{Fig:AfterglowSpectra}). We wish to evaluate if a blast wave with energy $10^{51}$ erg can reproduce this characteristic. We assume that the outflow can be described by a self-similar solution at 3 days. This is supported in the case of a jetted outflow due to the lack of a jet break in the radio light curve \citep{Toma2007}. 

The code used is described in detail in \citet{PeerWaxman2005}. It accounts for cyclosynchrotron emission and absorption, pair production and annihilation, and direct and inverse Compton processes including full Klein-Nishina corrections. 
The simulations also consider the external photon field from the rising supernova, to account for the effects of inverse (external) Compton scattering on the electron cooling. The code assumes one-zone emission in spherical symmetry.\footnote{The external, supernova photon field travels in the same direction as the jet. We include the de-boosting of the energy density and temperature of the supernova photons but the effect of inverse Compton scattering will be slightly overestimated in our analysis due to the assumption of spherical symmetry. However, this will not affect our conclusion. Indeed, weaker inverse Compton scattering would lead to a slight increase of the emission in the radio band.} 

Some additional modifications have been made to the code to be able to treat the problem at hand (we give a brief overview here, more details can be found in Appendix \ref{App:Code}). The Lorentz factor of the outflow is calculated from the blast wave kinetic energy $E_\textrm{k}$, the circumburst density $n_\textrm{cbm}$, and the radius $r$, following the Blandford-McKee solution in the relativistic case and the Sedov-Taylor solution in the nonrelativistic case. 
The two cases are smoothly connected with an interpolation between the two regimes. It is necessary to capture the trans-relativistic and nonrelativistic regimes, considering that the initial outflow velocity most likely was only mildly relativistic \citep{Campana2006, Soderberg2006Nature, Fan2006, Ghisellini2007, Toma2007}. 
Furthermore, the code allows the electrons to be injected in the downstream with a distribution consisting of both a NT and a thermal population, containing a fraction $\xi_a$ and $(1-\xi_a)$ of the particles respectively. 
As mentioned in the previous section, we assume no separation between the thermal and NT components in accordance with recent PIC simulations. 

The number fraction of accelerated electrons $\xi_a$ and the fraction of the downstream internal energy given to electrons $\epsilon_e$ are both input parameters of the code. Once set, they determine the temperature and shape of the injected electron distribution. Note that we define $\epsilon_e$ as the energy fraction received by \textit{all} electrons, i.e., 
\begin{equation}
    \epsilon_e = \epsilon_e^\textrm{th} + \epsilon_e^\textrm{NT},
\end{equation}
where $\epsilon_e^\textrm{th}$ and $\epsilon_e^\textrm{NT}$ are the internal energy fractions received by the thermal and NT electron population respectively. This is in contrast to the common definition of $\epsilon_e$ in the literature, where it most often refers solely to the energy fraction received by the NT electrons. However, this usually follows from assuming $\xi_a = 1$, which would be unrealistic for nonrelativistic and trans-relativistic shocks.

Other input parameters besides $E_\textrm{k}$, $n_\textrm{cbm}$, and $r$ are the electron injection index $p$ and the fraction of internal energy given to magnetic fields $\epsilon_B$. As previously mentioned, we fix $E_\textrm{k} = 10^{51}$ erg, as this is the minimum energy required to supply the UHECR flux. The radius is set so that we get an observed time of $\sim$ 3 days. The electron index is set to $p =2.1$, as suggested by the afterglow X-ray light curve \citep{Soderberg2006Nature, Fan2006, Toma2007}. We keep the density fixed at $n_\textrm{cbm} = 100$ cm$^{-3}$ as in Section \ref{Sec:Prompt}. We note that according to the scheme of \citet{Eichler2005}, the density should be increased as $n' \rightarrow n/\xi_a$ to retrieve similar afterglow characteristics. However, this would imply a circumburst density of $10^4$ cm$^{-3}$, much higher than what is commonly found in the literature ($n_\textrm{cbm} \sim 10^{-1}\textrm{--}10^2$ cm$^{-3}$ \citep{PanaitescuKumar2002}). Nevertheless, we checked the results for $n_\textrm{cbm} = 10^4$ cm$^{-3}$ and found that the conclusion remains the same. 

\subsection{Results}\label{Sec:ResultsAG}
\begin{figure}
\includegraphics[width=1\columnwidth]{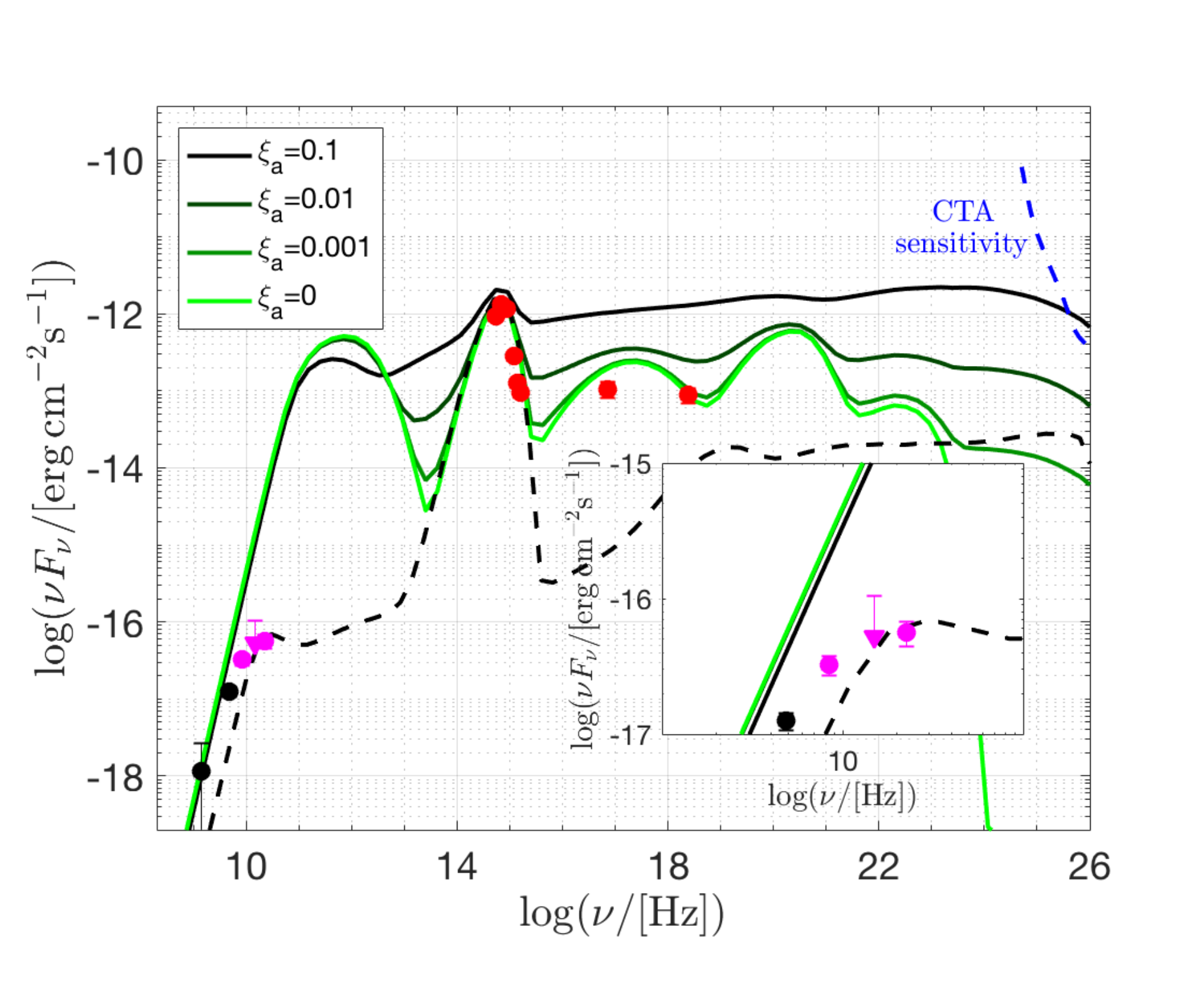}\hfill
    \caption{Examples of generated spectra for the forward shock emission at $\sim$ 3 days for different values of $\xi_a$ as indicated in the plot, using our code described in Subsection \ref{Sec:MethodologyAG}. Solid lines show spectra from a blast wave with the required $E_\textrm{k} = 10^{51}~$erg, while the dashed line shows a spectrum with $E_\textrm{k} = 10^{49}~$erg (for $\xi_a = 0.1$.) At low energies the spectrum is fully absorbed. The first bump at $\sim 10^{12}$ Hz is emission from the thermal electrons and the second bump at $\sim 10^{15}$ Hz is the rising supernova. At higher frequencies, the spectrum is dominated by synchrotron emission from the NT electrons and inverse Compton upscattering of the supernova photons by the thermal electrons. The latter is clearly visible in the $\xi_a = 10^{-3}$ case. While the optical-UV and X-ray data can be fit by decreasing $\xi_a$, the radio data cannot. The radio data is not overshone however, if the energy of the blast wave is decreased. Inset shows a zoom-in on the radio band. Black, magenta, and red data points are taken from \citet{Kaneko2007}, \citet{Soderberg2006Nature}, and  \citet{Campana2006} respectively. Error bars are given as 1$\sigma$ and upper limit (inverted triangle) is 3$\sigma$. The CTA design sensitivity is taken from \citet{CTAConsortium2011}. Other parameters used are $p = 2.1$, $n_\textrm{cbm} = 100$ cm$^{-3}$, $\epsilon_e = 0.1$, and $\epsilon_B = 10^{-3}$.}
    \label{Fig:AfterglowSpectra}
\end{figure}

In Figure \ref{Fig:AfterglowSpectra}, we show examples of our generated spectra for the forward shock at 3 days, assuming $\epsilon_e = 10^{-1}$ and $\epsilon_B = 10^{-3}$ for different values of $\xi_a$. The radio data are taken from \citet{Soderberg2006Nature} and \citet{Kaneko2007}. The optical data for the rising supernova and X-ray data are from \citet{Campana2006}. 
To get the X-ray data per frequency, we used the integrated XRT flux and $p=2.1$. 
From the figure, it is evident that the radio data point at 22.5 GHz is very constraining. Decreasing $\xi_a$ can make the emission compatible with the optical and X-ray data but it cannot be made consistent with the radio data. The reason for this is that the radio emission is dominated by the thermal electrons. Afterglow radio data is therefore essential in the study of UHECRs from GRB afterglows. In the figure, we include an example of a spectrum with the same parameters and $\xi_a = 0.1$ (dashed line), but the energy in the blast wave has been reduced from $10^{51}~$erg to $10^{49}~$erg. In this case, the spectrum does not overshoot the radio data.

\begin{figure}
\includegraphics[width=1\columnwidth]{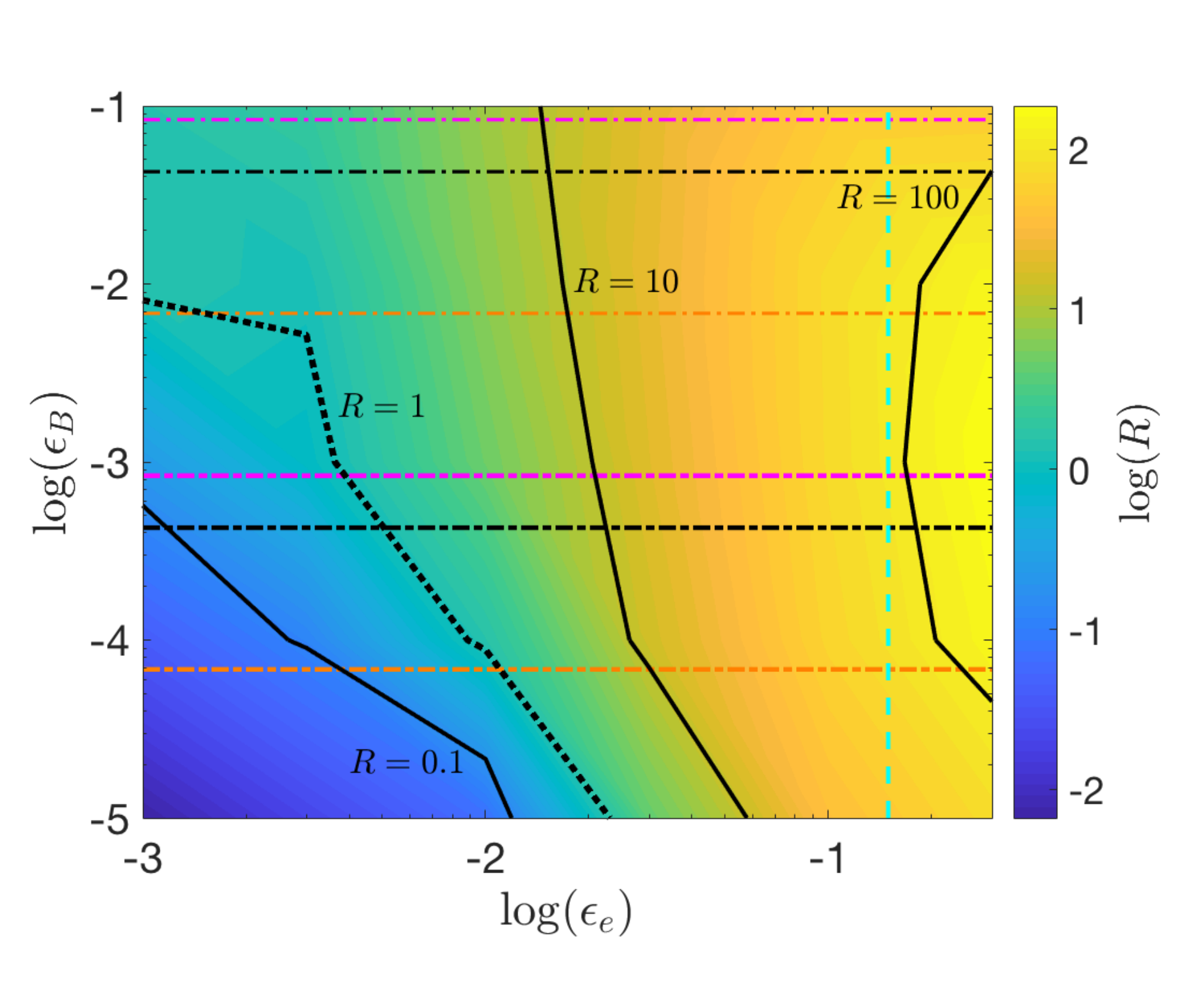}\hfill
    
    \caption{Color map showing $\log(R)$ as a function of $\epsilon_e$ and $\epsilon_B$, where $R$ is the ratio of the generated spectra to the 22.5 GHz data including the 1$\sigma$ uncertainty at 3 days (Equation \eqref{eq:Ratio}). Thick dotted, black line shows $R=1$, everything above overshoots the radio data at 1$\sigma$ confidence level. Solid black lines show other integer values of $\log(R)$ as indicated in the plot. Vertical dashed, cyan line corresponds to the value of $\epsilon_e$ found in PIC simulations of mildly relativistic shocks \citep{Crumley2019}. This value of $\epsilon_e$ would overshoot the radio data by more than an order of magnitude. The thick (thin) horizontal dotted-dashed lines show minimum values of $\epsilon_B$, required to accelerate UHECRs for $\eta = 1$ ($\eta = 0.1$) as given in Equation \eqref{eq:eB}. Black corresponds to $10^{20}$ eV for iron, while pink and orange correspond to the best fit rigidity as found by \citet{PierreAugerICRC2017}, and \citet{Heinze2019} respectively. Parameters used are $E_\textrm{k} = 10^{51}$ erg, $p = 2.1$, $n_\textrm{cbm} = 100$ cm$^{-3}$, and $\xi_a = 10^{-2}$.}
    \label{Fig:ee_eB_grid}
\end{figure}

Having determined that $\xi_a$ has little effect in the radio band, we fix $\xi_a = 10^{-2}$ and let $\epsilon_e$ and $\epsilon_B$ be free parameters to see what values are required to be consistent with the radio. The results are shown in Figure \ref{Fig:ee_eB_grid}. The color gradient shows the ratio of the generated spectra to the most constraining radio data point at 22.5 GHz, including the 1$\sigma$ error  provided in \cite{Soderberg2006Nature} as
\begin{equation}\label{eq:Ratio}
	R \equiv \frac{F_\textrm{radio}^\textrm{code}}{F^\textrm{measured}_\textrm{radio} + F^\textrm{error}_\textrm{radio}}.
\end{equation}
Thus, everything above $R=1$ overshoots the radio data at 1$\sigma$ confidence level. The figure shows that $\epsilon_e < 10^{-2}$ is necessary for the radio data to be consistent.


\subsection{Discussion}\label{Sec:DiscussionAfterglow}
Overlaid on the spectra in Figure \ref{Fig:AfterglowSpectra} is the sensitivity curve of the Cherenkov Telescope Array (CTA), taken for 50 hr exposure from \citet{CTAConsortium2011}. From the figure one can see that in the case of $\xi_a = 0.1$, the very high-energy gamma-ray emission is bright enough to be detectable by CTA. This means that in future GRB 060218-like events, CTA will be able to probe at least part of the possible parameter space. We note that this is sensitive to the electron index $p$.

From Figure \ref{Fig:ee_eB_grid}, it is evident that $\epsilon_e < 10^{-2}$ is required to not overshoot the radio data. Such small values of $\epsilon_e$ are not supported by the latest PIC simulations. In PIC simulations of mildly relativistic shocks, \citet{Crumley2019} found that $\epsilon_e^\textrm{NT}$ can indeed be very small. However, they also found that the electrons not accelerated into a power law thermalize at a temperature $T_e \sim 0.23 ~T_i$, where $T_i$ is the ion temperature in the downstream. In PIC simulations of nonrelativistic shocks, \citet{Bohdan2020arXiv} found similar values of $T_e \sim 0.15~T_i$ while \citet{Park2015} found even more effective thermalization with $T_e \sim T_i$. This heating via plasma processes of the thermal electrons requires $\epsilon_e$ to be large. In Figure \ref{Fig:ee_eB_grid}, we show $\epsilon_e = 0.15$ as a dashed vertical, cyan line, representing the value found by \citet{Crumley2019} ($\epsilon_e \sim 0.1 \textrm{--}0.2$). This value of $\epsilon_e$ would overshoot that radio data by $1\textrm{--}2$ orders of magnitude. Observations of supernova remnants also find the ratio $T_e/T_i$ to be between 0.03 and 1, albeit the velocities in those cases are much lower \citep{Ghavamian2013, Vink2015}. 

Very small values of $\epsilon_B$ help alleviate the constraint on $\epsilon_e$. However, small values of $\epsilon_B$ are problematic in the sense of UHECR acceleration. The dotted-dashed horizontal lines shown in Figure \ref{Fig:ee_eB_grid}, correspond to minimum values of $\epsilon_B$ required for UHECR acceleration. They are shown for different values of the best fit rigidities, similar to in Figure \ref{Fig:Robustness}. The lower limit on $\epsilon_B$ is calculated using $t'_{\textrm{acc},j} < t'_\textrm{ad}$ as 
\begin{equation}\label{eq:eB}
    \epsilon_B = \frac{B'^2}{8\pi u'_\textrm{int}} > \frac{1}{8\pi u'_\textrm{int}}\left(\frac{E}{\eta Z_j e r}\right)^2,
\end{equation}
where $u'_\textrm{int}$ is the internal energy of the downstream. 
The lines are plotted for completely stripped iron using acceleration efficiency $\eta = 0.1$ (thin) and $\eta = 1$ (thick).

It is possible that the magnetization decreases with distance to the shock front \citep[e.g.,][]{Lemoine2013}, something which is not considered in the present work. Inspection of Figure \ref{Fig:ee_eB_grid} shows that while $\epsilon_e \gtrsim 10^{-2}$, the radio flux is only weakly dependent on $\epsilon_B$. Therefore, for these values of $\epsilon_e$, our conclusion would not change if a varying magnetization was included. As previously mentioned, $\epsilon_e\sim 0.1\textrm{--}0.2$ is suggested by PIC simulations \citep{SironiSpitkovsky2011,Park2015,Crumley2019,Bohdan2020arXiv}.

The constraints shown in Figure \ref{Fig:ee_eB_grid} are valid for the microphysical parameters at the forward shock. In principle, the reverse shock could be different from the forward shock, which means that $\epsilon_B$ could be sufficiently small in the forward shock as to not overshine the radio, while large enough at the reverse shock for UHECR acceleration to be possible. As both the reverse and the forward shocks are trans-relativistic, this might imply that $\epsilon_B^\textrm{RS} \sim \epsilon_B^\textrm{FS}$ but this also depends on the magnetization around the two shocks, which can be different. 
One can estimate $\epsilon_B^\textrm{RS}$ as follows. Assuming cosmic ray acceleration occurs during the first shell crossing of the shock, the shell crossing radius is estimated as $R_x = [3E_\textrm{tot}\Delta/(4\pi n_\textrm{cbm}m_p c^2)]^{1/4} = 1.8\times10^{16}~\textrm{cm} ~ E_{\textrm{tot},51}^{1/4} ~ n_{\textrm{cmb},2}^{-1/4}$, where $\Delta = ct_\textrm{dur}$ is the shell width and $t_\textrm{dur}$ is the central engine duration that we set equal to $T_{90}$ \citep{NakarPiran2004}. With an initial Lorentz factor of $\Gamma = 10$, the ratio of the ejecta density $n_\textrm{ej}$ to the circumburst density is $n_\textrm{ej}/n_\textrm{cbm} > 1$. Thus, the Lorentz factor of the downstream of the reverse shock as seen from a laboratory observer can be estimated as $\Gamma_x \approx (\Gamma/2)^{1/2}(n_\textrm{ej}/n_\textrm{cbm})^{1/4} = 9.0 ~ E_{\textrm{tot},51}^{1/8} ~ n_{\textrm{cbm},2}^{-1/8}$, where $n_\textrm{ej} = E_\textrm{tot}/(4\pi m_p c^2 \Gamma^2 \Delta R_x^2)$ has been used \citep{PanaitescuKumar2004B,ZhangMurase2019}.\footnote{This approximation for $\Gamma_x$ is not very accurate in this case but has been used to show the parameter dependence. A more precise value is given by $\Gamma_x = \Gamma/(1+2\Gamma\sqrt{n_\textrm{cbm}/n_\textrm{ism}})
^{1/2} = 6.7$.} The magnetic field is calculated as $B'_x = (32\pi \epsilon_B^\textrm{RS} n_\textrm{cbm} m_p c^2 \Gamma_x^2)^{1/2}$. The value of $\epsilon_B^\textrm{RS}$ sufficient to accelerate UHECR can be found by comparing $B'_x$ to the the required magnetic field $B'_\textrm{req} = E/(\eta Z_j e R_x)$. The condition $B'_x/B'_\textrm{req} > 1$ translates to (considering iron)
\begin{equation}
    \epsilon_B^\textrm{RS} > 4.2\times 10^{-2} ~ E_{20}^{2} ~\eta_{-1}^2  ~ E_{\textrm{tot},51}^{-3/4} ~ n_{\textrm{cbm},2}^{-1/4}.
\end{equation}
Thus, with microphysical parameters $\epsilon_e^\textrm{FS} \sim 10^{-2}$, $\epsilon_B^\textrm{FS} \sim 10^{-4}$, and $\epsilon_B^\textrm{RS}\sim 5\times10^{-2}$ ($\epsilon_B^\textrm{RS}\sim 5\times10^{-4}$) for $\eta = 0.1$ ($\eta = 1$), UHECR acceleration at the reverse shock is not ruled out by our analysis.
If $\epsilon_B^\textrm{RS} \gg \epsilon_B^\textrm{FS}$, then one must make sure that the emission from the reverse shock is consistent with the radio and optical data as well, which is not considered in the current study. 

In this section we have shown, through analysis of the radio afterglow, that GRB 060218 is unlikely to have had a blast wave energy $E_\textrm{k} \gtrsim 10^{51}$ erg. As this is the energy necessary to supply the observed UHECR flux as argued in Section \ref{Sec:Energetics}, this result disfavors the afterglows of GRB 060218-like event to be the main source of UHECRs. Furthermore, this result constraints prompt models of UHECR acceleration as well. We find that in order to be compatible with the radio data at 3 days while simultaneously having sufficient energy available for the prompt phase, the energy in the prompt phase has to be roughly an order of magnitude higher than the blast wave kinetic energy. A prompt model that can accommodate UHECR acceleration therefore requires that $\gtrsim 90$\% of the energy escapes the system as cosmic rays, neutrinos, or radiation before the deceleration.


The spread in 8.5 GHz luminosities at 3 days for the llGRB sample in \citet{Margutti2013} is slightly less than an order of magnitude, with GRB 060218 at the lower end. One can naively assume that the spread at 22.5 GHz is roughly comparable to that at 8.5 GHz. With this assumption, the other llGRBs would have to fall somewhere between the $R=1$ and $R=10$ lines in Figure \ref{Fig:ee_eB_grid}, to be compatible with their respective radio data. This still requires $\epsilon_e$ to be smaller than the value found in the latest PIC simulations by at least a factor of a few. This is a hint that our conclusions can be applied to the current sample of llGRBs. 

The radio light curve of GRB 060218 does not exhibit any signs of a jet break. This led several authors to argue that the outflow of this burst was not collimated \citep{Soderberg2006Nature,Fan2006}. Alternatively, it was argued by \citet{Toma2007} that a jet break had already occurred before the radio observations began. In our analysis, we have assumed spherical symmetry of the outflow. 
We note here that the time evolution of the radio flux will be different after a jet break compared to a spherical outflow. 

How the radio flux evolves after the jet break depends on the relative positions of the injection frequency $\nu_\textrm{m}$ and the self-absorption frequency $\nu_\textrm{a}$ with respect to the radio frequency of interest $\nu_\textrm{radio} = 22.5~$GHz.\footnote{The spectrum suggests that the cooling break frequency $\nu_\textrm{c} \gg \nu_\textrm{radio}$ and can therefore be ignored here \citep{Soderberg2006Nature, Fan2006, Toma2007}.} 
If $\nu_\textrm{radio} < \nu_\textrm{a}$, the radio flux is either unaffected or increasing \citep{PanaitescuKumar2004}. This is because the decrease in flux due to the jet sideways expansion is counteracted by the increase in flux as the radio band becomes less and less absorbed. 

In our case, the radio band is dominated by the thermal emission, which is at $\sim\nu_\textrm{m}$ (see Appendix \ref{App:Code}). A major difference in radio flux between the spherical outflow and the jet break scenarios would occur if a spherical outflow predicts $\nu_\textrm{radio} \lesssim \nu_\textrm{m}$ while the jet break scenario results in $\nu_\textrm{m} \ll \nu_\textrm{radio}$, as the thermal emission will be well below the radio band in the latter case. The injection frequency evolves as $\nu_\textrm{m} \propto t^{-3/2}$ in the spherical outflow compared to $\nu_\textrm{m} \propto t^{-2}$ after the jet break \citep{PanaitescuKumar2004}.
According to \citet{Toma2007}, the jet break occurred at $2\times 10^4$ seconds after trigger, roughly a decade earlier in time compared to the radio observations at 3 days. This means that if a jet break occurred, our assumption of spherical symmetry overestimates $\nu_\textrm{m}$ by a factor of $\sim 3$.
Inspection of Figure \ref{Fig:AfterglowSpectra} shows that a frequency shift of the thermal emission by a factor 3 only decreases the flux by a factor of a few. We therefore conclude that relaxing our assumption of spherical symmetry to include a possible jet break would not substantially alter our results. 
This once again shows the importance of early afterglow radio data, with which an early jet break could have been observed.

In this paper, we have limited ourselves to studying the fastest part of the outflows of llGRBs, responsible for the radio emission in the first few days. However, there are several other transients reported or suggested in the literature where our methodology could be applied. For instance, llGRBs are associated with energetic Type Ib/c supernovae. The associated supernova ejecta, expanding at $\lesssim 0.1~c$, carries most of explosion energy but is too slow to accelerate cosmic rays to more than a few $10^{18}~$eV. If, however, the velocity distribution of the ejecta is continuous between the fast llGRB ejecta and the slow supernova ejecta, then a significant amount of the energy will be contained in a trans-relativistic outflow with speeds of $0.3\textrm{--}0.5~c$ \citep{Margutti2013}. At this TRSN component, UHECR acceleration could occur. Additionally, the TRSN would also carry enough energy to supply the observed UHECR flux \citep{ZhangMurase2019}. A potential trans-relativistic component is difficult to constrain as it starts to decelerate much later than the mildly relativistic ejecta (weeks to months after trigger). If late-time radio detections from future llGRB events can be obtained, a possible TRSN component could be constrained using the methodology of this section. 

An additional important application is emission from the reverse shock, where UHECR acceleration could occur \citep{WaxmanBahcall2000,PeerWaxman2005b,ZhangMurase2019}. 
Thermal synchrotron emission from electrons accelerated in the downstream of the reverse shock should exist if $\xi_a \ll 1$.
Another example is the trans-relativistic SN 2009bb that was discovered through radio observations \citep{Soderberg2010}. Although no $\gamma$-ray emission was detected in this event, the radio emission suggested the outflow was at least mildly relativistic. The supernova had similar radio characteristics as previously detected llGRBs and \citet{Soderberg2010} estimated the event rate of SN 2009bb-like events to be comparable to the rate of llGRBs.
Lastly, we mention the peculiar fast blue optical transients (FBOTs) AT 2018cow \citep{Margutti2019}, ZTF18abvkwla \citep{Ho2020}, and CSS161010 \citep{Coppejans2020}, whose late-time radio luminosities are comparable to that of llGRBs. The current methodology can either be used to study the connection between these trans-relativistic transients and UHECRs or as a diagnostic of their total kinetic energy.

\section{Conclusion}
\label{Sec:Conclusion}
In this paper, we have studied whether the mildly relativistic outflows of llGRBs can be the main source of UHECRs, using the canonical low-luminosity GRB 060218 as a proxy. 
Our investigation has focused on the inevitable radiation from the electrons in the UHECR acceleration region. We have proposed that synchrotron emission from thermal electrons serve as a powerful probe of the physics of mildly relativistic shocks and cosmic ray acceleration. In particular, searching for thermal synchrotron emission at the radio band gives us constraints on the kinetic energy of the mildly relativistic ejecta, by which the possibility of UHECR acceleration in llGRBs and TRSNe can be critically tested. 
Together with the approach used in \citet{Samuelsson2019}, it can efficiently constrain the UHECR acceleration site. As the electron synchrotron spectrum carries additional information to other messengers, such as neutrinos and hadronic gamma-rays, it is also a very useful tool in multi-messenger modeling of UHECR sources.

This paper extended on the work of \citet{Samuelsson2019} by not only studying the prompt phase, but also the possibility of UHECR acceleration in the afterglow phase. Another difference of this work compared to \citet{Samuelsson2019} is that by using the specific low-luminosity GRB 060218 as a representative source, we could resort to direct measurements instead of generic assumptions for the llGRB fluxes. As a consequence, the results presented here are stronger and more comprehensive. 

For the prompt phase, we got requirements on the comoving magnetic field at the source by comparing the UHECR acceleration time scale to typical energy loss time scales. 
Given the magnetic field, we characterized the synchrotron flux emitted by the coaccelerated electrons assuming these were instantaneously injected into a power law distribution \citep{BlumenthalGould1970, Sari1998}. This mostly resulted in a fast-cooling spectrum with a low-energy index of $-0.5$. For our fiducial parameters given in Table \ref{Tab:NumericalValues}, we found that the flux would be orders of magnitude higher than observed values, especially in the optical-UV band (Figure \ref{Fig:FluxLimits}). As such, high optical fluxes could be a tell-tale signature of successful UHECR acceleration during the prompt phase. 

For our optimistic parameters, the acceleration efficiency was increased to $\eta= 1$, the fraction of internal energy given to NT electrons was decreased to $\epsilon_e^\textrm{NT} = 5\times10^{-4}$, and the number fraction of accelerated electrons $\xi_a$ was made a free parameter, while the other parameters remained unchanged (Figure \ref{Fig:Robustness}). 
For $\Gamma = 3$, we found no viable solution for UHECR acceleration even with these optimistic parameters. This value of $\Gamma$ is consistent with the small values of $\Gamma \lesssim 5$ argued for by a majority of previous studies \citep[e.g.,][]{Campana2006, Soderberg2006Nature, Fan2006, Ghisellini2007, Toma2007, Waxman2007}.
Within the current uncertainties, for $\Gamma = 10$ the UHECR observations could only be explained in a fine-tuned region if $\xi_a \lesssim 10^{-2}$ and the acceleration took place at $\gtrsim 5\times 10^{15}$ cm, or if $\xi_a < 2\times 10^{-4}$, where an iron energy of $4\times 10^{19}$ eV could be reached. For $\Gamma = 30$ or larger and $\xi_a = 10^{-2}$ or less, we could not constrain UHECR acceleration. 
However, this solution relied on the shocks being mildly relativistic. If $\epsilon_e^\textrm{NT}$ was fixed at 0.1 as often assumed in the case of canonical high-luminosity GRBs, this solution disappeared even for $\eta = 1$ and lower values of $\xi_a$.

For the prompt phase, we argued that because GRB 060218 had higher observed prompt optical luminosity compared to other llGRBs, the constraints for the whole population might actually be stronger than those presented here. While we have not aimed to reproduce the observed spectrum, previous studies have assumed the phenomenological shape of the observed spectrum to compute the cooling of UHECRs \citep{Murase2006, Murase2008, Liu2011, Zhang2018}. Specifically, they used a harder low-energy slope that is by construction already consistent with the optical-UV constraint. Other mechanisms beyond the synchrotron model used in this paper would be necessary to accommodate such a spectrum and we leave it for future work to see what restrictions can be put on such models. 
The conclusions from this part of the paper were valid for radii that satisfied $t_\textrm{v} < T_{90}/2$ only, as discussed in Section \ref{Sec:SubFlux}.

We then considered UHECR acceleration on the reverse (or forward) shock of the afterglow phase, as considered by \citet{ZhangMurase2019}. To match the observed UHECR flux on Earth, the kinetic energy of the mildly relativistic blast wave had to be large ($\geq 10^{51}$ erg), much larger than suggested by previous afterglow studies of GRB 060218. This indicated that $\xi_a$ had to be small ($\sim 10^{-2}$).
Small values of $\xi_a$ mean that the bulk number of electrons ($1-\xi_a$) are thermal and their contribution to the emission is non-negligible. Using the numerical code of \citet{PeerWaxman2005}, modified to capture the problem at hand, we found that these thermal electrons mostly radiate in the radio band. Afterglow radio data a few days after trigger is therefore essential in constraining the parameters and, consequentially, in discerning if the afterglows of llGRBs can be the main sources of UHECRs or not. 
Furthermore, we showed that CTA will be able to probe part of the parameter space in future GRB 060218-like events. 

Using the necessary blast wave energy of $10^{51}$ erg for the mildly relativistic ejecta, we got constraints on the microphysical parameters $\epsilon_e$ and $\epsilon_B$. The results showed that $\epsilon_e < 10^{-2}$ is required to be consistent with the radio data at 3 days. Here, $\epsilon_e$ is the energy fraction shared by \textit{all} electrons. In PIC simulations of mildly relativistic shocks, \citet{Crumley2019} found that $\epsilon_e \sim 0.1\textrm{--}0.2$, which would overshoot the 1$\sigma$ upper limit of the radio data by more than an order of magnitude (Figure \ref{Fig:ee_eB_grid}). 
This result gives interesting constraints on prompt models as well. To have sufficient energy available in the prompt phase while simultaneously being consistent with the afterglow radio data implies that $\gtrsim 90$\% of the initial energy escaped the system as cosmic rays, neutrinos, or radiation before deceleration. 

We stress that the results of this work are general and useful whether llGRBs are the sources of UHECRs or not.  
The implications for UHECRs are summarized as follows. In the prompt phase, UHECRs acceleration is largely excluded mainly because fast-cooling spectra overshoot the optical limit, although models that can produce the observed harder low-energy slope are not constrained. 
In the afterglow scenario, we have shown that thermal synchrotron emission gives stringent constraints when UHECRs are accelerated by a mildly relativistic component with $\Gamma \beta \sim 2\textrm{--} 10$.
The early radio data can be explained with a smaller amount of kinetic energy ($E_\textrm{k}\sim 10^{49}~$erg), in which the energy budget of UHECRs would be insufficient even if they can be accelerated. 
However, the results do not exclude that they are dominantly accelerated at slower trans-relativistic components of the outflow whose deceleration occurs at much later times (weeks or months). 
Nevertheless, the strategy proposed here is general, and future applications to other llGRBs, their reverse shock emission, and TRSNe will enable us to get important insights into the physics of UHECR acceleration and related shock physics.

\acknowledgments
We thank B. Theodore Zhang and the anonymous referee for fruitful discussions. We acknowledge support from the Swedish National Space Agency (196/16) and the Swedish Research Council (Vetenskapsr\aa det, 2018-03513). Financial support is also acknowledged from the Swedish Foundation for international Cooperation in Research and Higher Education (STINT). D.B. is supported by the Deutsche Forschungsgemeinschaft (SFB 1258). F.R. is supported by the G\"oran Gustafsson Foundation for Research in Natural Sciences and Medicine. A.P. is supported by the European Research Council via ERC consolidating grant \#773062 (acronym O.M.J.). K.M. is supported by the Alfred P. Sloan Foundation, NSF Grant No. AST-1908689, and KAKENHI No. 20H01901.

\begin{table}
\begin{center}
\caption{Numerical values used unless otherwise stated.}
  \begin{tabular}{ l | l | l }\label{Tab:NumericalValues}
  	Quantity & Symbol & Value used \\ \hline
  	\hline
  	Redshift & $z$ & $0.033$ \\ 
  	Prompt optical flux & $F_{\nu_\textrm{opt}}$ & 0.55 mJy\\
    Prompt X-ray flux & $F_{\nu_X}$ & 0.1 mJy \\
  	Optical band energy & $h\nu_{\rm opt}$ & 3 eV \\
    X-ray band energy & $h\nu_\textrm{X}$ & 5 keV\\
    Typical observed photon energy & $\left< \varepsilon \right>$ & 5 keV\\ 
    Total energy & $E_\textrm{tot}$ & $10^{51}$ erg \\
    Total luminosity & $L_\textrm{tot}$ & $4.8\times10^{47}$ erg s$^{-1}$\\ 
    Radiation luminosity & $L_\gamma$ & $3.0\times10^{46}$ erg s$^{-1}$\\ 
    Photohadronic cross section & $\sigma_{p\gamma}$ & $10^{-28}$ cm$^2$\\ 
    Acceleration efficiency & $\eta$ & $0.1$ \\ 
  	NT electron energy fraction & $\epsilon_e^\textrm{NT}$ & 0.1 \\
  	Electron injection index & $p$ & 2.5 \\
    Electron acceleration fraction & $\xi_a$ & 1 \\  
    Constant in $\gamma'_{\rm m}$ & $a$ & 1
  \end{tabular}
\end{center}
\end{table}

\bibliographystyle{mnras}
\bibliography{References} 


\appendix
\section{Synchrotron self-Compton scattering during the prompt phase}\label{App:InverseCompton}
To accommodate to XRT and BAT data, the SSC emission during the prompt phase of GRB 060218 cannot be large \citep{Ghisellini2007}. Therefore, we neglected SSC emission in the calculation of the prompt spectra in Section \ref{Sec:Prompt}. In this appendix, we discuss how SSC would affect our results. 

Assuming we are in the fast-cooling regime, which is the most relevant case when considering UHECR acceleration, and assuming Compton scattering to be Klein-Nishina suppressed after one scattering, the Compton Y-parameter is given by \citep[e.g.,][]{Sari1996}
\begin{equation}
\begin{split}
	Y = 
	\begin{cases}
		\sqrt{U'_e/U'_B} \qquad & \textrm{if} ~ U'_e \gg U'_B, \\[3mm]
		 U'_e/U'_B & \textrm{if} ~ U'_e \ll U'_B.
	\end{cases}
\end{split}
\end{equation}
The magnetic field energy density $U'_B$ is given by
\begin{equation}
	U'_B = \frac{B'^2}{8\pi},
\end{equation}
and the electron energy density $U'_e$ can be estimated as 
\begin{equation}
	U'_e = \frac{\epsilon_e L_\textrm{tot}}{4\pi r^2 c \Gamma^2}.
\end{equation}
With the same parameter values as those used in Section \ref{Sec:SubFlux}, we obtain
\begin{equation}
	 \frac{U'_e}{U'_B} = 6.7 ~ \epsilon_{e,-1}~ L_{\textrm{tot}, 48} ~ r_{14}^{-2} ~ \Gamma_1^{-2} ~ (B'_{3})^{-2},
\end{equation}
which gives 
\begin{equation}\label{eq:Y-value}
	 Y \approx  2.6 ~ \epsilon_{e,-1}^{1/2} ~ L_{\textrm{tot}, 48}^{1/2} ~ r_{14}^{-1} ~ \Gamma_1^{-1} ~ (B'_{3})^{-1}.
\end{equation}
From \eqref{eq:Y-value}, one sees that the Compton $Y$-parameter is around unity. High values of the magnetic field necessary to obtain the highest energy UHECRs, and small values of $\epsilon_e$ as used for our optimistic parameters, both help to suppress $Y$. For our fiducial and our optimistic parameter sets, acceleration of iron to $10^{20}~$eV always results in $Y \lesssim ~$a few, regardless of $\Gamma$ and $r$. Therefore, we conclude that our results would only slightly change if SSC was included and our conclusion would remain. 

For small values of the magnetic field sufficient to accelerate the lower energy UHECRs, the $Y$-parameter can become large. The effect of SSC on the synchrotron spectrum is to lower the cooling frequency $\nu_\textrm{c}$ by a factor of $Y^2$, which in turn decreases $\nu_\textrm{SSA}$. The maximum flux given in Equation \eqref{eq:F_max} is independent of $Y$. The flux in the lower energy bands therefore increases, as the maximum synchrotron flux is shifted toward lower frequencies. The flux in the higher energy bands decreases for the same reason. Specifically, the flux around the $\nu F_\nu$-peak decreases by a factor $1+Y$. Therefore, the constraints on UHECR acceleration may become more or less severe, depending on the values of $\nu_\textrm{opt}$ and $\nu_\textrm{X}$ compared to $\nu_\textrm{c}$, $\nu_\textrm{m}$, and $\nu_\textrm{SSA}$. Furthermore, a large $Y$-parameter implies a significant inverse Compton component to be present in the prompt spectrum, something which can be effectively constrained by the XRT and BAT data. Indeed, the lack of a clear inverse Compton signature in the observed spectrum can be used to further constrain the allowed magnetic field strength and hence, the possibility of UHECR acceleration. However, given that the $Y$-parameter is close to unity for the highest UHECR energies as mentioned in the discussion above, incorporating this additional constraint from SSC would not qualitatively alter the conclusion.

\section{Details of the code}\label{App:Code}
In this section, we present the modifications made to the code, compared to the description of \citet{PeerWaxman2005}. First, we explain how the code treats the nonrelativistic, trans-relativistic, and ultra-relativistic cases self-consistently, and after we show how the NT plus thermal electron distribution is calculated. 

The Lorentz factor of the shock $\Gamma_\textrm{s}$ is tailored to smoothly transition between the relativistic and the nonrelativistic regimes. It is calculated as follows. We introduce the dimensionless quantity $\zeta$ as 
\begin{equation}
	\zeta = \left(\frac{E_\textrm{k}}{n_\textrm{cbm}m_pc^2r^3}\right)^{1/2}.
\end{equation}
Then, $\Gamma_\textrm{s}$ is given by
\begin{equation}\label{eq:Gamma_s}
	\Gamma_\textrm{s} = \sqrt{\left(\zeta \left \{ \frac{2}{5} + \left(\sqrt{\frac{17}{8\pi}} - \frac{2}{5} \right)\frac{\zeta^2}{1+\zeta^2}\right \} \right)^2 +1}.
\end{equation}
When $\zeta$ is large, the kinetic energy of the blast wave is much larger than the swept up rest-mass energy and the Blandford-McKee solution is valid. For $\zeta \gg 1$, equation \eqref{eq:Gamma_s} simplifies to 
\begin{equation}
	\Gamma_\textrm{s} \approx \sqrt{\left(\sqrt{\frac{17}{8\pi}} \zeta \right)^2 +1} \approx \left(\frac{17E_\textrm{k}}{8\pi n_\textrm{cbm}m_pc^2r^3}\right)^{1/2},
\end{equation}
which is the correct expression when $\Gamma_\textrm{s} \gg 1$. When $\zeta$ is small, we wish to retrieve the Sedov-Taylor solution. In this case, equation \eqref{eq:Gamma_s} simplifies to 
\begin{equation}
	\Gamma_\textrm{s} \approx \sqrt{\left(\frac{2}{5} \zeta \right)^2 +1}.
\end{equation}
Thus, we get the velocity in units of speed of light as 
\begin{equation}
	\beta_\textrm{s} = \sqrt{1-\Gamma_\textrm{s}^{-2}} \approx \frac{2}{5}\left(\frac{E_\textrm{k}}{n_\textrm{cbm}m_pc^2r^3}\right)^{1/2}.
\end{equation}
This is the correct solution, assuming the proportionality constant in $r(t)$ in the Sedov-Taylor solution is of order unity.

The internal energy density (excluding rest-mass energy) in the downstream is given by \citep{BlandfordMcKee1976}
\begin{equation}\label{eq:u_int}
	u'_\textrm{int} = (\Gamma-1)\frac{\hat{\gamma}\Gamma + 1}{\hat{\gamma}-1}n_\textrm{cbm}m_pc^2,
\end{equation}
where $\Gamma$ is the Lorentz factor of the downstream region and $\hat{\gamma}$ is the adiabatic index of the downstream region. To calculate $\Gamma$, we need to know $\hat{\gamma}$, which is in itself a function of $\Gamma$ \citep{BlandfordMcKee1976}. The value of $\hat{\gamma}$ will be somewhere between $4/3$ (ultra-relativistic) and $5/3$ (nonrelativistic). This problem is solved iteratively, using the approximation for the adiabatic index given in \citet{Service1986}, accurate to one in $10^{5}$. Whenever $\epsilon_e$ and $\epsilon_B$ are given, they refer to fractions of the internal energy density as given in equation \eqref{eq:u_int}. 
%
%
%
%
%
 
The NT electrons population will be accelerated from the thermal bulk, as suggested by PIC simulations \citep{SironiSpitkovsky2011,Park2015,Crumley2019,Bohdan2019}. This means that the temperature of the thermal population will be correlated to the injection energy of the NT population. In practice, this is implemented by enforcing  $\theta' = \gamma'_\textrm{m}(\beta'_\textrm{m})^2/(1+(\beta'_\textrm{m})^2)$, where $\theta'$ is the comoving temperature of the thermal electrons in units of electron rest mass and $\beta'_\textrm{m} = \sqrt{1-(\gamma'_\textrm{m})^{-2}}$. With this choice, we get $\theta' = \gamma'_\textrm{m}/2$ for $\gamma'_\textrm{m}\beta'_\textrm{m} \gg 1$ and $\theta' = (\beta'_\textrm{m})^2$ for $\gamma'_\textrm{m}\beta'_\textrm{m} \ll 1$.

\label{lastpage}
\end{document}